\documentclass[fleqn,usenatbib]{mnras}

\usepackage{newtxtext,newtxmath}
\usepackage[T1]{fontenc}

\DeclareRobustCommand{\VAN}[3]{#2}
\let\VANthebibliography\thebibliography
\def\thebibliography{\DeclareRobustCommand{\VAN}[3]{##3}\VANthebibliography}

\usepackage{graphicx}	% Including figure files
\usepackage{amsmath}	% Advanced maths commands
\usepackage[usenames,dvipsnames]{color}
\usepackage{color}

\usepackage{ulem}

\newcommand{\tpm}{t$_\text{PM}$}

\newcommand{\QPM}{Q$_\text{PM}$}
\newcommand{\QCTRL}{Q$_\text{CTRL}$}

\newcommand{\eag}{EAGLE}
\newcommand{\tng}{TNG100-1}
\newcommand{\ill}{Illustris-1}

\defcitealias{Quai2021}{Q21} %% use \citetalias{Quai2021}

\title[The interconnection between galaxy mergers, AGN and quenching]{The interconnection between galaxy mergers, AGN activity and rapid quenching of star formation in simulated post-merger galaxies.}

\author[S. Quai et al.]{
Salvatore Quai,$^{1,2,3}$\thanks{E-mail: salvatore.quai@unibo.it}, Shoshannah Byrne-Mamahit,$^{1}$ Sara L. Ellison,$^{1}$ David R. Patton,$^{4}$ Maan H. Hani$^{5}$
\\
$^{1}$Department of Physics and Astronomy, University of Victoria, 3800 Finnerty Rd, Victoria, BC V8P 5C2, Canada\\
$^{2}$Dipartimento di Fisica e Astronomia ``Augusto Righi'', Universit\'{a} degli Studi di Bologna, Via Gobetti 93/2, I-40129 Bologna, Italy\\
$^{3}$INAF-Osservatorio di Astrofisica e Scienze dello Spazio di Bologna, Via Gobetti 93/3, I-40129 Bologna, Italy\\
$^{4}$ Department of Physics and Astronomy, Trent University, 1600 West Bank Drive, Peterborough, ON K9L 0G2, Canada\\
$^{5}$ Department of Physics and Astronomy, McMaster University Hamilton ON L8S 4M1, Canada\\
}

\date{Accepted XXX. Received YYY; in original form ZZZ}

\pubyear{2015}

\begin{document}
\label{firstpage}
\pagerange{\pageref{firstpage}--\pageref{lastpage}}
\maketitle

\begin{abstract}
We investigate the role of galaxy mergers on supermassive black hole (SMBH) accretion and star formation quenching in three state-of-the-art cosmological simulations with contrasting physics models: EAGLE, Illustris and IllustrisTNG.   We find that recently coalesced 'post-mergers' in all three simulations have elevated SMBH accretion rates by factors of $\sim2-5$.  However, rapid (within $500$ Myr of coalescence) quenching of star formation is rare, with incidence rates of $0.4\%$ in Illustris, $4.5\%$ in EAGLE and $10\%$ in IllustrisTNG.  The rarity of quenching in post-mergers results from substantial gas reservoirs that remain intact after the merger.  The post-mergers that do successfully quench tend to be those that had both low pre-merger gas fractions as well as those that experience the largest gas losses.  Although rare, the recently quenched fraction of post-mergers is still elevated compared to a control sample of non-mergers by factors of two in IllustrisTNG and 11 in EAGLE.  Conversely, quenching is \textit{rarer} in Illustris post-mergers than in their control.  Recent observational results by Ellison et al. have found rapid quenching to be at least $30$ times more common in post-mergers, a significantly higher excess than found in any of the simulations.  Our results therefore indicate that whilst merger-induced SMBH accretion is a widespread prediction of the simulation, its link to quenching depends sensitively on the physics models, and that none of the subgrid models of the simulations studied here can fully capture the connection between mergers and rapid quenching seen in observations.
\end{abstract}


%%%%%%%%%%%%%%%%%%%%%%%%%%%%%%%%%%%%%%%%%%%%%%%%%%

%%%%%%%%%%%%%%%%% BODY OF PAPER %%%%%%%%%%%%%%%%%%

%%%%%%%%%%%%%%%%%%%%%%%%%%%%%%%%%%%%%%%%%%%%%
%%%%%%%%%%%%%%%%%%%%%%%%%%%%%%%%%%%%%%%%%%%%%
%%%%%%%%%%%%%%%%%%%%%%%%%%%%%%%%%%%%%%%%%%%%%
\section{Introduction}

%\noindent- Theory: mergers trigger inflows of gas that fuel enhanced central SFR and BH accretion. Then, AGN feedback and quenching.
%
%\noindent- Observations: moderate SFR enhanement; central metal dilution (not so central, though). 
%Still debating whether:
%1) mergers are the main trigger of AGNs,
%2) AGN-driven quenching.
%
%\noindent- Improvement of subgrid models: state-of-the-art cosmological simulations now predict:
%
%* moderate SFR enhancement (more in accordance with observations) - Hani+2020
%
%* moderate BH accretion enhancement - Shoshannah+in prep.
%
%* quenching is rare in post merger galaxies - Quai+2021
%
%
%\noindent - This paper: 
%
%* 
%
%* comparison BH accretion in post-merger galaxies from different sub-grid models.
%
%* comparison of impact of AGN feedback on quenching SFR. 

Mergers are amongst the most catastrophic events that occur in galaxies.
Early works with N-body simulations on idealised mergers showed how tidal forces dramatically distort and deform merger remnants \citep[][]{Toomre1972, White1978, Roos1979}. 
Both theory \citep[e.g.,][]{Villumsen1982, DiMatteo2007, Cox2008, Lotz2008, Moreno2015} and observations \citep[e.g.,][]{HernandezToledo2005, Patton2005, DePropris2007, Casteels2014, Patton2016} indicate that the ultimate morphological outcome of the interaction is driven by orbital properties and mass ratio of the progenitors, and by physical properties of the involved galaxies.

Furthermore, simulations have long predicted that galaxy mergers induce gravitational instabilities in the interstellar medium (ISM) due to a redistribution of angular momentum, thus driving gas towards the inner part of galaxies \citep[e.g.,][]{Hernquist1989, Barnes1996, Mihos1996, Blumenthal2018}.
Theory predicts that such strong gas inflows from the outer disk to the inner kpcs should induce an enhancement in central star formation rates \citep[e.g.,][]{Cox2008, DiMatteo2008, Hopkins2008, Sparre2016},and a dilution in central gas-phase metallicity \citep[e.g.,][]{Montuori2010, Torrey2012, Bustamante2018}.
Observational studies of galaxy-galaxy mergers give support to the theoretical framework. 
Merging galaxies show enhanced SFRs \citep[e.g.][]{Li2008, Ellison2008, Patton2013, Knapen2015, Cao2016}, and central metallicity dilution \citep[e.g.,][]{Ellison2013, Gronnow2015, Thorp2019, Bustamante2020}.

Following the pioneering work of \cite{Springel2005}, galaxy formation models generally invoke active galactic nuclei (AGN) feedback mechanisms to efficiently quench star formation in massive galaxies, especially as a consequence of a merger \citep[e.g.,][]{DiMatteo2005, Hopkins2008}.
Indeed, during galaxy-galaxy interactions, theory predicts that part of the in-falling gas should feed the central super-massive black hole (SMBH), thus leading to rapid black hole growth and triggering AGN feedback with the release of energy \citep[e.g.,][]{LyndenBell1969, Cattaneo2009, Capelo2015}.
A fraction of the energy ejected from the AGN feedback, in turn, couples with the surrounding interstellar medium and drives out a significant amount of gas from the galaxy \citep[e.g.,][]{Silk1998,  Murray2005, Fabian2006, Cattaneo2009}.

However, in the last decade there has been an intense debate in the community about the role of mergers in triggering AGN activity. 
Many observational studies did not find evidence of merger-driven AGN activity
\citep[e.g.,][]{Cisternas2011, Schawinski2011, Kocevski2012, Villforth2014, Hewlett2017, Shah2020, Lambrides2021}, while many others
have found robust proofs for an AGN excess in interacting galaxies \citep[e.g.,][]{Alonso2007, Ellison2011, Bessiere2012, Ellison2013b, Kocevski2015, Hewlett2017, Marian2020, Pierce2022}.
It is possible divergent results arise from different experimental set up, including (not exhaustively) mixed definitions of AGN, mixed low and high redshifts selection, lack of matched controls, AGN fraction in mergers versus excess of mergers amongst AGN, and/or different wavelength range involved \citep[e.g., a larger importance of mergers in mid-IR AGNs than in optically selected ones,][]{Satyapal2014, Goulding2018, Ellison2019, Gao2020}. 
However, \citet{Ellison2019} ruled out the possibility that 
the tension could be due to different methodologies, at least for the low redshift infrared selected AGN.
%divergent results arise from %different experimental set up  %\citep[e.g., AGN fraction in %mergers versus excess of mergers %amongst AGN,][]{Ellison2019} %\textcolor{red}{(I don't think %this is a good reference for this %statement, since Sara finds an %excess in both methods %demonstrating that the tension %does not exist due to methodology %at least for the low redshift IR %selected AGN)}, and/or different %wavelength range involved %\citep[e.g., a larger importance %of mergers in mid-IR AGNs than in %optically selected ones,][]{Satyapal2014, Goulding2018, Ellison2019, Gao2020}  
Furthermore, outcomes from cosmological simulations suggest that the majority of low-redshift AGN should not reside in recent mergers \citep[][]{Steinborn2018,McAlpine2020,ByrneMamahit2022}. 
However, even though most AGN are not triggered by mergers, mergers can still lead to AGN.  
For example, in \citet{ByrneMamahit2022} we demonstrated that post-mergers from the IllustrisTNG simulation \citep{Marinacci2018, Naiman2018, Nelson2018, Pillepich2018b, Springel2018} host an excess of AGN and that the majority of post-mergers have above average SMBH accretion rates, consistent with the prediction of enhanced AGN activity.
However, there is no consensus about either the physical processes governing feedback mechanisms associated with AGNs (thermal energy, kinetic energy, a combination of the two), nor its extent (range, scales involved, strength). 
Different models implement different prescriptions for describing AGN feedback and the fraction of energy released, often leading to a dissimilar impact on galaxy evolution \citep[e.g.,][]{Somerville2015, Schaye2015, Weinberger2017}. 
It is therefore important to investigate whether the results of \cite{ByrneMamahit2022} are robust to variations in AGN feedback models.
 The first major goal of this paper is to rigorously compare the AGN-driven merging scenario in three state-of-the-art cosmological simulations: \eag{} \citep{Schaye2015, Crain2015, McAlpine2016}, \ill{} \citep{Vogelsberger2014}, and \tng{}.

Beyond the mere triggering of nuclear accretion, theoretical predictions of merger-driven AGN activity have long suggested that galaxy mergers might be the leading mechanism for rapid galaxy quenching \citep[e.g.,][]{DiMatteo2005, Springel2005a, Springel2005c, Croton2006, Hopkins2008, Somerville2008}. 
Effective mechanical AGN feedback triggered by gas with low angular momentum which is accreting onto the central SMBH could potentially drive out the gas from the galaxy and heat the circumgalactic medium (CGM), thus halting star formation and hampering the
replenishment of the galactic gas supply. 

Many previous simulations that linked mergers to AGN activity were able to efficiently remove gas from the galaxy reservoir, leading to a prompt quenching shortly after the coalescence phase \citep[e.g.,][]{Springel2005c, Bower2006, Khalatyan2008, Hopkins2008}. Such early works tended to implement very aggressive AGN feedback prescriptions, thus provoking an immediate evacuation of the gas from most post-coalescence galaxies. 
Recent results also support a pivotal role of mergers to halt star formation. For instance, studies based on the so called ``genetic modification approach'' \citep{Roth2016}, that are capable of altering the accretion history of a galaxy by generating sets of controlled
numerical realisations in a fully cosmological context, have demonstrated 
a causal connection between mergers and quenching \citep[e.g.,][]{Pontzen2017, Sanchez2020, Davies2022}. 
The genetic approach grants the advantage  of controlling the main parameters involved in the galaxy interaction, thus allowing the search for causal connections between the various physical processes involved. However, currently the results are limited to a restricted number of objects covering a small dynamical range of masses, orbits, mass ratios and other physical properties that might be at play during mergers.  

Many other recent works based on the statistical analysis of cosmological hydrodynamical simulations, which include complete samples of mergers covering a realistic range of parameters such as galaxy morphology, gas fraction, and orbital geometry, have instead challenged the merger-driven quenching scenario \citep[e.g.][]{Weinberger2018, Correa2019, RodriguezMontero2019, Quai2021, Pathak2021}. In \citet{Quai2021}, we quantified that only $\sim5-10$ percent of post-mergers selected from an unbiased IllustrisTNG300-1 galaxy sample quench as a direct consequence of mergers. 
In the current work, our second major goal is therefore to test whether the lack of merger-driven quenching is ubiquitous in the three aforementioned cosmological simulations and how their different prescriptions for AGN activity reflect upon the outcome of the quenched fraction in the global post-merger population. 

A pivotal point that warrants examination is whether or not the merger-driven quenching scenario is in accordance with observational results. 
Proving a causal connection between mergers and quenching with observations is challenging.
Among the known populations of recently quenched galaxies \citep[e.g.,][]{Citro2017, Quai2018, Quai2019}, the most promising quenched galaxies that can be directly linked to galaxy-galaxy interactions are those known as `post-starburst', also called `E + A' or `K + A' galaxies. 
Post-starburst galaxies are systems showing spectroscopical evidence of a sharp decline of star formation following an intense star-burst episode that occurred between 0.5-1 Gyr ago \citep[e.g.,][]{Dressler1983, Couch1987, Wild2010, Pawlik2019, Wild2020}. Morphological studies show that post-starbursts often exhibit morphological disturbance and tidal features \citep[e.g.,][]{Oegerle1991, Zabludoff1996, Yang2008, Pawlik2018, Sazonova2021, Wilkinson2022}, all characteristics suggesting a merger-driven quenching scenario.
To date, most of the observational studies of the merger-driven quenching scenario have focused on determining the frequency of mergers in various post-starburst populations selected via a plethora of methods, finding contrasting results ranging from small fractions \citep[$\lesssim10\%$,][]{Blake2004, Setton2022}, intermediate fractions \citep[$20-40\%$,][]{Wilkinson2022}, and high fractions \citep[up to $60\%$,][]{Pracy2009}. However, for the purpose of this paper, we are interested in the complementary question of determining what is the fraction of rapidly quenched galaxies in an unbiased sample of recent post-merger galaxies. 
Thanks to the first statistically robust large sample of pure recent post-merger galaxies selected by \citet{Bickley2022}, \citet{Ellison2022} could recently address the latter question, finding a strong excess of quenching in post-mergers. 
Quantitatively, \citet{Ellison2022} demonstrated that a rapid truncation of star formation is $30-60$ times more probable in post-coalescence galaxies than expected in a control sample of non-interacting galaxies, demonstrating an incommensurable dissimilarity with the quenching excess of \citet{Quai2021}. 
The third major goal of this paper is therefore to compare for the first time observational results of merger triggered quenching \citep[from][]{Ellison2022} with the predictions from the earlier mentioned cosmological simulations, to gain insights into the interconnection between galaxy mergers and star formation quenching.

The paper is organised as follows: in Section~\ref{sec:data}, we briefly introduce the cosmological simulations used in this work, as well as the methodology to select post-mergers and a control sample of non-interacting galaxies. 
In Section~\ref{sec:results} we present the results. 
In detail, in Section~\ref{sec:accretion} we quantify the enhancement of AGN activity in star-forming post-merger galaxies; in Section~\ref{sec:fractions} we measure the impact of mergers on quenching star-formation; in Section~\ref{sec:observations}, we compare the quenching excess in simulated post-mergers with that from observations. 
Finally, we discuss the results in Section~\ref{sec:discussion}, and we summarise our work in Section~\ref{sec:conclusions}.

%However, \citet{Wilkinson2022} demonstrated that numbers of factors can negatively impact the merger fraction of the post-starburst population, showing that $70\%$ of recent mergers in IllustrisTNG would not be selected as post-mergers with any of the currently available identification metrics, suggesting that many more post-starburst galaxies could have experienced a recent merger event.

%%%%%%%%%%%%%%%%%%%%%%%%%%%%%%%%%%%%%%%%%%%%%
%%%%%%%%%%%%%%%%%%%%%%%%%%%%%%%%%%%%%%%%%%%%%
%%%%%%%%%%%%%%%%%%%%%%%%%%%%%%%%%%%%%%%%%%%%%
\section{Data and methods}
\label{sec:data}
Black hole accretion and AGN feedback operate on scales below the simulation resolution, and thus require subgrid prescriptions for modelling the physical processes involved.
Distinct cosmological simulations might formulate different sub-grid recipes, responding to (1) appropriate calibrations of the simulations via tuning of physical parameters required to reproduce fundamental observed relation scales, and (2) updated models following a continuous improvement of our knowledge on the physical processes.
As mentioned in the Introduction, most of the modern hydro-dynamical simulations use AGN feedback to halt star formation in massive galaxies, therefore, the choice of different models for regulating black hole accretion and AGN feedback may have important consequences on the merger-driven quenching scenario we aim to address. 
We compare the response of three state-of the art cosmological simulations (i.e., \eag{}, \ill{}, and \tng{}) to describe the interconnection between merger events, black hole activity, and quenching of star formation.
Even though the three simulations span a similar volume (cubes of sides around $100$ cMpc), and all of them assume black hole accretion to be proportional to the spherically symmetric Bondi–Hoyle–Lyttleton accretion rate \citep{Hoyle1940, Bondi1944, Bondi1952}, they opt for subtly different physical adjustments and implement distinct AGN feedback models, that might result in a diverse response to galaxy merger events. 
In this section, we give a brief overview of the simulations we use in this work and their sub-grid models governing black hole activity (see a brief summary of the black hole model prescriptions in Table~\ref{tab:sample_prop}). 
Readers interested in details about different models for AGN feedback and consequent implications on SMBH properties and galaxy evolution are referred to \cite{Habouzit2021, Habouzit2022}.

%%%%%%%%%%%%%%%%%%%%%%%%%%%%%%%%%%%%%%%%%%%%%
%%%%%%%%%%%%%%%%%%%%%%%%%%%%%%%%%%%%%%%%%%%%%
\subsection{Cosmological simulations}
%%%%%%%%%%%%%%%%%%%%%%%%%%%%%%%%%%%%%%%%%%%%%
%%%%%%%%%%%%%%%%%%%%%%%%%%%%%%%%%%%%%%%%%%%%%

\begin{table*}
%\begin{minipage}{\linewidth}
\centering
\caption{A brief summary of the prescriptions for the models of black hole accretion \protect\citep[i.e., note that the three simulations employ accretion rates proportional to the Bondi–Hoyle–Lyttleton accretion rate][]{Hoyle1940, Bondi1944, Bondi1952}, and AGN feedback in the simulations \eag{}, \ill{}, and \tng{}.}
\label{tab:sample_prop}
\begin{tabular}{|p{0.1\linewidth}|p{0.25\linewidth}|p{0.25\linewidth}|p{0.25\linewidth}|}
& \eag{} & \ill{} & \tng{} \\
 \hline 
 \hline 
BH accretion & The model uses the angular momentum of the surrounding gas to reduce BH accretion rates in small galaxies.
& The accretion rate depends on the gas properties at the location of the BH    (susceptible to local stochasticity) & A Kernel-weighted accretion rate over $256$ neighbouring  cells. The accretion rate correlates with the gas properties in the central regions.
\\
\hline
AGN feedback &
One-mode:
\begin{itemize}
\item thermal, with a net efficiency $0.015$;
\item the energy is released isotropically.
\end{itemize}
&
Two-mode (both thermal):
\begin{itemize}
\item At $f_\text{Edd}>=0.05$,  thermal energy is released with a net coupling efficiency $0.05$.
\item At $f_\text{Edd}<0.05$ thermal energy is released in hot bubbles within a radius of $\sim100$ kpc from the BHs. 
\end{itemize}
&
Two-mode:
\begin{itemize}
\item At high accretion rates, a thermal ``quasar'' mode, similar to Illustris.
\item At low accretion rates, a kinetic ``radio'' mode, with directional injection of momentum into surrounding ISM, oriented in a random direction (isotropic feedback over several events). 
\end{itemize}
\\
\hline
Note: & The energy is released following a minimum heating temperature criterion: BHs store feedback energy until a threshold amount. This prevents numerical losses and overcooling.
 & The `hot bubble' model impacts more on the host halo (M/M$_\text{halo}$ too high).
& TNG uses a switch for low- and high-accretion rates with a dependence  on the BH mass (low-accretion mechanical mode effective at M$_\text{BH} \geq 10^{8.2}$M$_\odot$).
\\
\hline
\end{tabular}		 
%\end{minipage}\hfill
\end{table*}

%%%%%%%%%%%%%%%%%%%%%%%%%%%%%%%%%%%%%%%%%%%%%
%%%%%%%%%%%%%%%%%%%%%%%%%%%%%%%%%%%%%%%%%%%%%
\subsubsection{\eag{}}
\label{sec:eagle}
The ``Evolution and Assembly of GaLaxies and their Environment'' \citep[\eag{},][]{Schaye2015, Crain2015, McAlpine2016} is a suite of hydrodynamical cosmological simulations spanning a wide range of numerical resolutions and physical models which are implemented using a bespoke version of the Smoothed Particle Hydrodynamics (SPH) code GADGET-3 \citep{Springel2005c}.
For this work, we use the largest run denoted Ref-L0100N1504, a cubic periodic volume of ($100$ cMpc)$^3$ with dark matter and stellar mass resolutions of $9.7\times10^6$M$_\odot$ and
$1.8\times10^6$M$_\odot$, respectively.

The \eag{} public catalogues contain only $29$ snapshots at redshift z$<20$, yelding a temporal resolution that is insufficient to cover the rapid changes a galaxy faces during mergers. We use instead a galaxy catalogue  (internal team use) with a much denser time grid, including $201$ snipshots at redshift z$<20$, $65$ of which at z$\leq1$ are useful for this work.
The cosmological parameters used in \eag{} are in accordance with \citet{PlanckCollaboration2014} which is given by a matter density $\Omega_\text{M,0} = 0.307$, baryon density $\Omega_\text{b,0} = 0.04825$, dark energy density $\Omega_{\Lambda,\text{0}} = 0.693$, a Hubble parameter $h = 0.6777$, and $\sigma_8 = 0.8288$.

Star formation is implemented via a sub-grid model described in \citet{Schaye2008} and based on a pressure-dependent relation that converts gas to stars following the Schmidt-Kennicutt law \citep{Kennicutt1998}.

Following the prescription in  \cite{Springel2005b}, black holes are seeded as collision-less sink particles with an initial mass of  $1.475\times10^5$ M$_\odot$ in the centres of black hole-free dark matter halos exceeding a threshold mass of $1.475\times10^{10}$ M$_\odot$. Subsequently, \eag{} black holes can grow either via (1) accretion of surrounding material or (2) black hole mergers. 
Black hole accretion is regulated following 
a modified Bondi-Hoyle rate \citep{Bondi1944} that includes a prescription that takes into account  the angular momentum of material nearby a black hole \citep[see][]{RosasGuevara2015, RosasGuevara2016}. The inclusion of the angular momentum of surrounding gas allows the material to first settle into an accretion disk, thus reducing black hole accretion rates in small galaxies compared to the original Bondi-Hoyle model. 
Finally, in \eag{} accretion rates are capped at the Eddington limit, to avoid reaching spuriously high accretion rate values. 

\eag{} uses single-mode AGN feedback \citep{Booth2009}, where energy is injected thermally into the surrounding ISM, with a fixed efficiency (i.e., a fraction of the accreted gas is stochastically converted in thermal energy and released in the region nearby a black hole with a net efficiency of $0.015$), independently from halo mass and accretion rate.
The thermal energy is released following a minimum heating temperature criterion, that is when the black hole has accreted enough mass that the equivalent energy released is capable of increasing the temperature of the surrounding gas particles, thus preventing numerical losses and overcooling.

%%%%%%%%%%%%%%%%%%%%%%%%%%%%%%%%%%%%%%%%%%%%%
%%%%%%%%%%%%%%%%%%%%%%%%%%%%%%%%%%%%%%%%%%%%%
\subsubsection{\ill{}}
\label{sec:illustris}

We identify post-mergers galaxies also in the Illustris cosmological simulation  \citep{Vogelsberger2013, Vogelsberger2014}.
In this work, we use the flagship \ill{} \citep{Nelson2015} run, that covers a comoving cubic volume of side 106.5 cMpc. 
\ill{} has a dark matter and baryonic mass resolutions of $6.3\times10^6$ M$_\odot$ and
$1.6\times10^6$ M$_\odot$, respectively.
The simulation runs from redshift $127$ to the present day ($50$ snapshots at redshift z$\leq1$)
using the AREPO moving-mesh code \citep{Springel2010, Pakmor2016}. The cosmological parameters used in IllustrisTNG are in accordance with WMAP 9 \citep{Hinshaw2013} which is given by a matter density $\Omega_\text{M,0} = 0.2726$, baryon density $\Omega_\text{b,0} = 0.0456$, dark energy density $\Omega_{\Lambda,\text{0}} = 0.7274$, and a Hubble parameter $h = 0.704$, and $\sigma_8 = 0.809$.

Star formation is implemented in a sub-grid model following the \citet{Springel2003} formalism.
Gas particles exceeding a hydrogen number density of $0.13$ cm$^{-3}$ are described by a ``star-forming'' effective equation of state, and their gas is converted to stars stochastically following the Schmidt-Kennicutt law \citep{Kennicutt1998} assuming a \citet{Chabrier2003} initial mass function \citep[see][for further details]{Nelson2015}.

In \ill{}, black hole seeding follows the prescription in \cite{Springel2005b}, where massive halos exceeding M$_\text{halo}=7.1\times10^{10}$ M$_\odot$ and devoid of central supermassive black holes are seeded with a black hole of initial mass M$_\text{BH}=1.42\times10^{5}$ M$\odot$. In Illustris, the accretion rate onto the black hole follows an Eddington-capped Bondi-Hoyle-Lyttleton prescription. 
The accretion rate scales with the local mass density of the gas particle located at the physical position of the black hole itself.
To compensate for the unresolved multiphase ISM \citep{Springel2005}, that produces a non-physical under-density in the region surrounding the central black holes, the Illustris model introduces a boost factor that increases the instantaneous accretion rate. 

Illustris uses a two-mode thermal AGN feedback, with both modes proportional to the amount of mass accumulated into the central black hole. 
At high accretion rates (i.e., at Eddington fractions  \emph{f}$_\text{Edd}=\dot{\text{M}}_\text{BH}/\dot{\text{M}}_\text{Edd}>0.05$), Illustris activates the so-called ``quasar'' mode, where the AGN deposits thermal energy in the region surrounding the black hole, with a net coupling efficiency of $0.05$.
Instead, at low accretion rates (i.e., at Eddington fractions  \emph{f}$_\text{Edd}<0.05$), Illustris switches to the ``radio'' mode AGN feedback. The recipe for the low accretion rate regime should mimic the influence of an AGN radio jet (not resolved in the simulation), with hot bubbles of thermal energy deposited in the CGM within a radius of $100$ kpc from the central black hole, coupling to the gas with a net efficiency of $0.35$. At high accretion rates, the feedback is implemented as a continuous injection  of thermal energy into the ISM, thus subject to numerical losses. 
Therefore, the AGN quench galaxies exclusively through the low accretion mode.

%%%%%%%%%%%%%%%%%%%%%%%%%%%%%%%%%%%%%%%%%%%%%
%%%%%%%%%%%%%%%%%%%%%%%%%%%%%%%%%%%%%%%%%%%%%
\subsubsection{Illustris\tng{}}
\label{sec:tng}
%The work we present here is primarily aimed at %quantifying the impact of galaxy mergers on star %formation quenching.
%We identify galaxy post-mergers in the %IllustrisTNG simulation suite
%\citep{Nelson2019} to study the relationship %between mergers and the interruption of star %formation within a cosmological framework.
The IllustrisTNG (TNG) project includes a suite of large-box magnetohydrodynamical cosmological simulations in a $\Lambda$CDM Universe
which provides a statistically robust sample of galaxies spanning a variety of galaxy properties (e.g., mass, environment, star formation rate). 
The simulations and physical model are introduced in detail in \citet{Marinacci2018, Naiman2018, Nelson2018, Pillepich2018b, Springel2018}.
IllustrisTNG (or TNG) is the descendant of the Illustris cosmological simulation with an improved physical models and numerical scheme.  
TNG also introduces a number of additional features to obtain a better agreement with observational results. 
In this paper, we focus on TNG100-1, the highest resolution run for the intermediate publicly released volume of side $110.7$ cMpc. 
TNG100-1 offers excellent statistics, whilst still guaranteeing adequate numerical resolution. TNG100-1 has a dark matter and stellar mass resolutions of $7.5\times10^6$ M$_\odot$ and
$1.4\times10^6$ M$_\odot$, respectively.
The simulation runs from redshift $127$ to the present day
using the AREPO moving-mesh code \citep{Springel2010, Pakmor2016}. The cosmological parameters used in IllustrisTNG are in accordance with \citet{PlanckCollaboration2016} which is given by a matter density $\Omega_\text{M,0} = 0.3089$, baryon density $\Omega_\text{b,0} = 0.0486$, dark energy density $\Omega_{\Lambda,\text{0}} = 0.6911$, and a Hubble parameter $h = 0.6774$. 

Star formation occurs in a pressurised, multi-phase interstellar medium following the \citet{Springel2003} formalism.
Gas particles whose density exceeds a threshold of $\sim0.1$ cm$^{-3}$ are ``star-forming'' and their gas is converted to stars stochastically following the Schmidt-Kennicutt law \citep{Kennicutt1998} assuming a \citet{Chabrier2003} initial mass function \citep[see][for further details]{Nelson2015, Pillepich2018a}.

Black holes are seeded with an initial mass of $1.18\times10^6$M$_\odot$ at the centres of the potential wells of halos exceeding a threshold mass of $7.38\times10^{10}$M$_\odot$.
Black holes can grow their mass either through (1)
accretion following a modify Bondi-Hoyle scheme, or (2) mergers with other black holes. The instantaneous accretion rate in IllustrisTNG does not depend upon the gas particle at the location of the black hole (as in Illustris). Instead, the rate scales following a kernel-weighted gas density over about $256$ neighbouring particles to reduce the dependence on the local gas mass fluctuations, and it is capped at the Eddington rate.

AGN feedback is directly related to the accretion rate onto the central black holes ($\dot{\text{E}} \propto \dot{\text{M}}_{\text{BH}}\text{c}^2$).
At high accretion rates (i.e., quasar mode
feedback), thermal energy is returned to the black hole's environment, whereas at low accretion rates (i.e., radio mode feedback, or kinetic feedback), energy accumulates until it reaches an energy threshold, then mechanical energy is instantaneously released along a random direction into the gas around the black hole \citep[see][for further details]{Weinberger2017}. As in Illustris, in IllustrisTNG the high accretion mode is subject to numerical losses. Therefore, quenching in massive galaxies can be obtained only through the kinetic mode at low accretion rates \citep[e.g.,][]{Terrazas2020}.
TNG uses a switch for activating either low- or high-accretion rates. The switch function depends on the black hole mass, with low-accretion mechanical mode becoming effective at M$_\text{BH} \geq 10^{8.2}$M$_\odot$.

%%%%%%%%%%%%%%%%%%%%%%%%%%%%%%%%%%%%%%%%%%%%%
%%%%%%%%%%%%%%%%%%%%%%%%%%%%%%%%%%%%%%%%%%%%%
%%%%%%%%%%%%%%%%%%%%%%%%%%%%%%%%%%%%%%%%%%%%%
\subsection{Methods}
\label{sec:method}

For the purpose of this paper, for each galaxy and for its descendants, we retrieve and use stellar, halo, gas and black hole masses, as well as instantaneous star formation rates. SMBH accretion rates are calculated by averaging the change in SMBH mass between snapshots over the duration of the snapshot. The mean snapshot intervals are $\sim 118$ Myrs, $\sim 156$ Myrs, and $\sim 159$ Myrs for \eag{}, \ill{}, and \tng{} respectively. 
\subsection{Star-forming post-mergers}
%%%%%%%%%%%%%%%%%%%%%%%%%%%%%%
We select and follow the evolution of galaxies from the three simulations' merger trees created using the {\scshape Sublink} code \citep{RodriguezGomez2015}. 
Galaxy mergers are defined as nodes in the {\scshape Sublink} merger trees \citep{RodriguezGomez2015}. 
Namely, we define a post-merger (or PM) in the snapshot immediately after the coalescence phase as the remnant of two interacting galaxies.
Following \citet{Hani2020} and \citet{Quai2021}, our post-merger sample is restricted to those satisfying the following criteria. 
First of all, post-mergers must be star-forming when first selected (in order that we can later observe them quenching). In practice, this is implemented by requiring SFRs higher than $-1\sigma$ from the star-forming main sequence best-fit in the given snapshot (i.e., at fixed redshift).
We only follow the evolution of post-mergers at redshift z $\leq 1$ (i.e., in the last $\sim8-9$ Gyr), in order to avoid the vigorous merger activity at higher redshift that would aggravate issues such as numerical stripping and subhalo switching \citep{RodriguezGomez2015}, and to grant a robust sample of long-term non-interacting galaxies that constitute the pool for selecting a suitable control population.
Moreover, we limit the analysis to galaxies more massive than M$_\ast \geq 10^{10}$ M$_\odot$, thus ensuring a complete sample of well-resolved post-mergers with a mass ratio (secondary/primary) larger than 1:10. 
We then apply a mass ratio cut, excluding mergers with a mass ratio (secondary/primary) $\mu < 0.1$.
This criterion is strictly connected with the previous one and prevents the analysis of remnants of mergers with a companion less massive than the reliable mass limit of the simulation. 

\begin{figure*}
\setlength\lineskip{-3.5pt}
\centering
\includegraphics[width=0.497\linewidth]{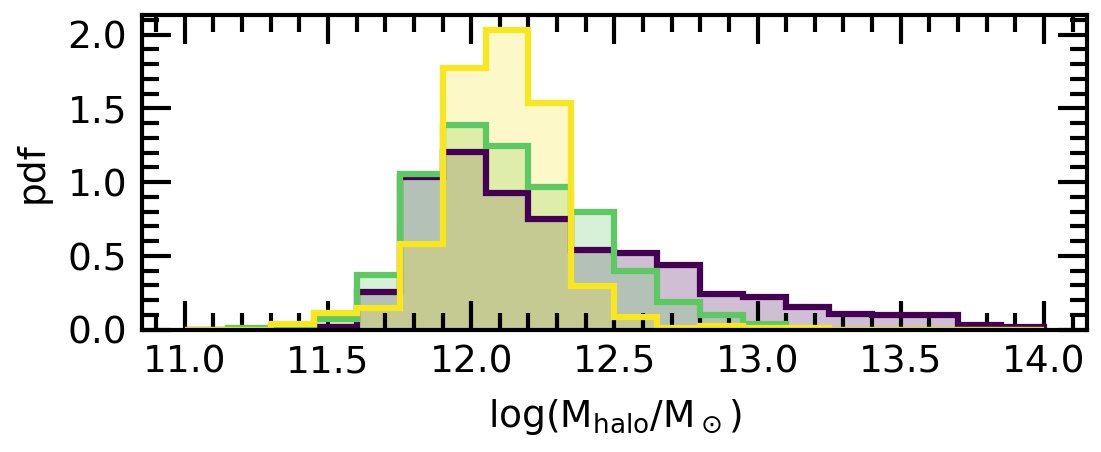}
\includegraphics[width=0.497\linewidth]{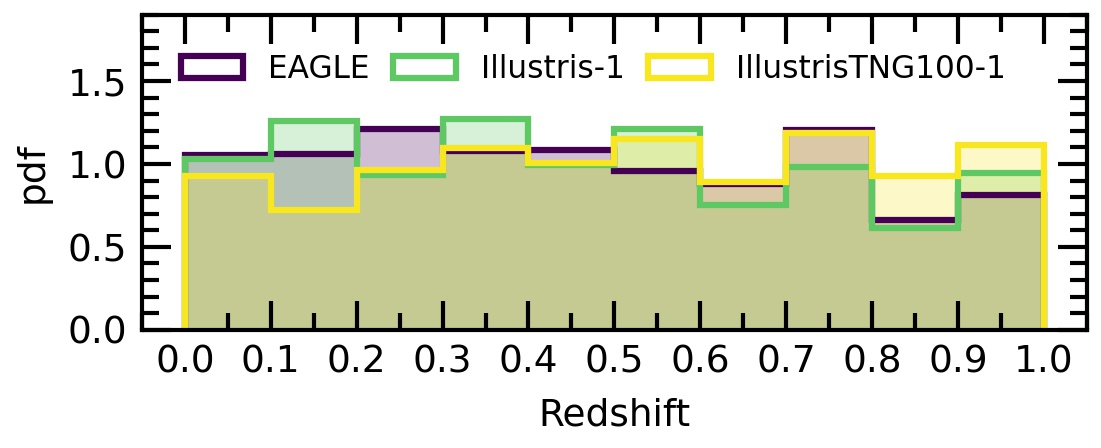}\\
\includegraphics[width=0.497\linewidth]{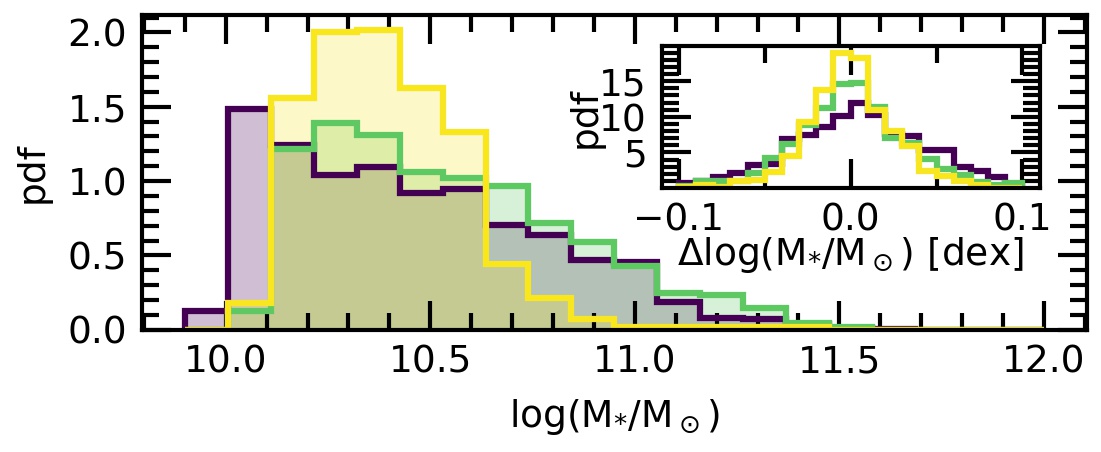}
\includegraphics[width=0.497\linewidth]{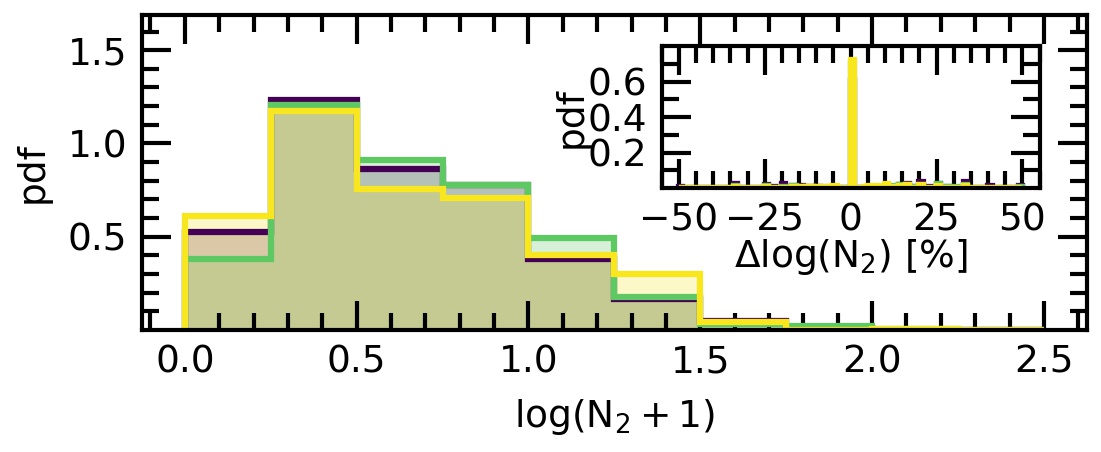}\\
\includegraphics[width=0.497\linewidth]{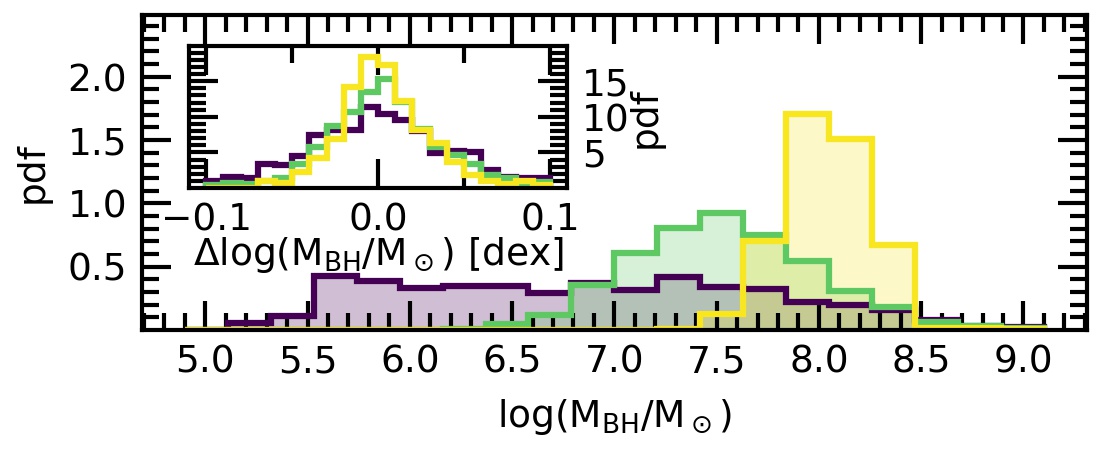}
\includegraphics[width=0.497\linewidth]{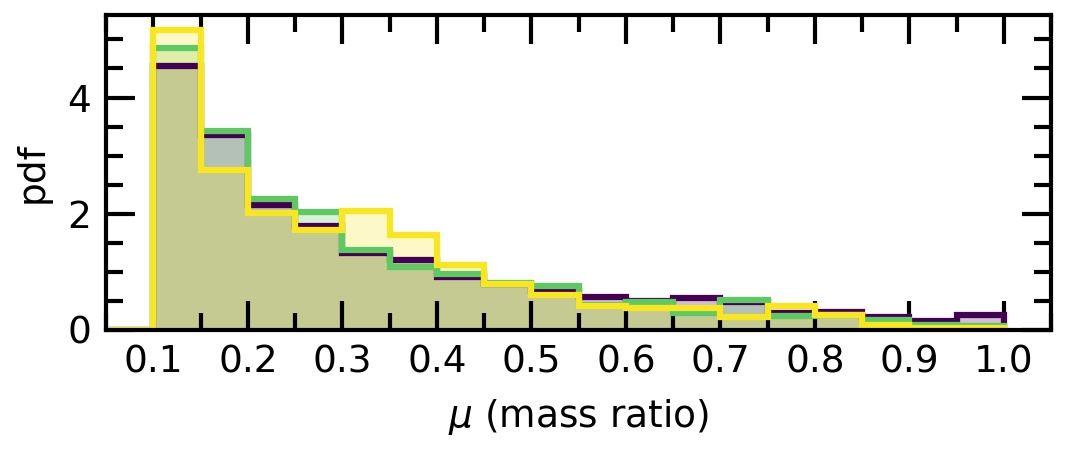}\\
\caption{Distribution of the main parameters of star-forming post-mergers selected from the \eag{} (purple), \ill{} (green), and \tng{} (yellow) simulations. In the left column, we show (from top to bottom) halo mass, stellar mass, and SMBH mass distributions, whilst in the right column, we show (from top to bottom) redshift, the number of neighbours within two Mpc (N$_2$), and mass ratio ($\mu$) distributions. 
The three insets in the stellar mass, SMBH mass, and N$_2$ panels represent the distribution of differences (i.e. post-mergers values minus control galaxy values) in the three matched parameters (plus redshift, that is matched de facto as well) following our matching scheme.}
\label{fig:distributions}
\end{figure*}

For \ill{} and \tng{} post-mergers, we further prevent numerical stripping and subhalo switching issues \citep{RodriguezGomez2015} by adopting the following procedure to estimate the mass ratio of the merger. 
We define the mass ratio ($\mu$) as the ``mean''  mass ratio value measured over maximum $10$ snapshots leading up to coalescence. We exclude the maximum and minimum mass ratios from the mean calculation to remove possible extreme variations which we found to be strongly related to the numerical stripping and subhalo switching effect, respectively. In \eag{}, instead, the mass ratio $\mu$ is from \citet{McAlpine2020} and it is computed when the in-falling galaxy onto the main progenitor had its maximum mass.
 
 Finally, for \ill{} and \tng{} post-mergers we also apply two additional environmental constraints aimed at avoiding post-mergers that are undergoing further close interactions.
 Following \citet{Patton2020}, we measure  r$_\text{sep}$ as:
 \begin{equation}
 \text{r}_\text{sep} = \frac{\text{r}}{\text{R}^\text{host}_\text{1/2} + \text{R}^\text{comp}_\text{1/2}},
 \end{equation}
where r is the 3D separation between the centres of the host (i.e. post-merger in our case) and its closest neighbour, and R$^\text{host}_\text{1/2}$ and R$^\text{comp}_\text{1/2}$ are the stellar half mass radii of the post-merger and the closest neighbour, respectively. 
\citet{Patton2020} showed that \tng{} galaxies with r$_\text{sep} < 2$ had stronger numerical mass stripping effects, hence we exclude them from the following analysis. 
The second environmental parameter is r$_1$, that is defined in \citet{Patton2020} as the 3D separation to the nearest neighbour with stellar mass above $10\%$ of the mass of the galaxy in question. We exclude post-mergers that have r$_1<100$ kpc, whose companion could be already interacting with the post-mergers.

Our selection criteria yield three samples at redshift z$\leq1$ of $1551$, $1082$, and $560$ star-forming post-mergers for \eag{}, \ill{}, and \tng{}, respectively. 
Despite the fact that the three simulations span similar comoving volumes and use similar cosmology, the number of star-forming post-mergers in \tng{} is remarkably smaller than in the other two simulations.
Figure~\ref{fig:distributions} shows the distribution in the main parameters (i.e., redshift, stellar mass, gas fraction (f$_\text{gas}$, i.e., the fraction of the total gas over the sum of stellar and gas mass), halo mass, SFR, SMBH mass, mass ratio $\mu$, and N$_2$) defining the \eag{}, \ill{}, and \tng{} post-merger populations.
The different number of \tng{}'s star-forming post-mergers compared to the other two simulations is explained by the difference in the halo and stellar mass distributions.  
In \tng{}, indeed, there are only a few star-forming post-mergers with M$_\text{halo}>10§^{12.2}$ M$_\odot$ and M$_\ast > 10§^{10.5}$ M$_\odot$, whilst the rest of \tng{} massive post-mergers are passively evolving.
The three simulations show significant differences in the black hole mass distributions, with \tng{} post-mergers harbouring already massive SMBH with masses around $10^{8}$ M$_\odot$, whilst \ill{} and \eag{} post-mergers show a shallower black hole mass distribution, with \eag{} having a considerable tail of galaxies with black holes less massive than $10^6$ M$_\odot$.  
Instead, the three simulations show a similar distribution in redshift, local density N$_2$ \citep[defined in ][as the number of neighbours within a radius of $2$ Mpc]{Patton2016}, and mass ratio.
About the latter parameter, we note that $50.2\%$, $52.6\%$  and $49.6\%$ of \eag{}, \ill{}, and \tng{} post-mergers are minor mergers with mass ratio $\mu<0.25$ (i.e., 1:4 mergers).
We address the impact of mass ratio on quenching star formation in Section~\ref{sec:analysis}.

\subsection{Statistical control sample}
\label{sec:control_sample}
We are interested in investigating the link between galaxy mergers, AGN activity, and the quenching of star formation. 
We quantify the impact of galaxy mergers on the quenching of star-formation using an observational approach that consists of identifying control galaxies that are matched to each post-merger galaxy in redshift, stellar mass, and environment \citep[e.g.,][]{Ellison2013, Patton2013}. 
In this section, we describe the steps of generating the control sample.

We implement an adaptation of the matching procedure used in \citet{Patton2016} and \citet{Patton2020} to select statistical controls (CTRL) for post-merger galaxies in our sample.   
For each star-forming post-merger, we define a star-forming control pool (i.e. non-interacting galaxies with SFR higher than $-1\sigma$ from the SFMS), at the same redshift (i.e. same snapshot/snipshot), with M$_\ast \geq 10^{10}$ M$_\odot$, with a relative separation from the nearest neighbour r$_\text{sep} \geq 2$, and r$_1\geq100$ kpc. We also exclude galaxies that have experienced a merger ($\mu \geq 0.1$) within 2 Gyr. 
%Moreover, we narrow the control sub-sample of each post-merger to the galaxies with a match in:
Then, for each post-merger in our sample, we identify the control galaxies that match the post-merger properties as follows:  
\begin{itemize}
\item log(M$_\ast$) within a tolerance of $0.1$ dex.

\item log(M$_\text{BH}$) within a tolerance of $0.1$ dex. 

% By matching in black hole mass, any dissimilarity in quenching fraction between the post-merger and the non-interacting control samples cannot be ascribed to a difference in AGN activity. ###I commented this out because I'm not sure how I feel about this statement since it contradicts the accretion rate excess that we are finding -- i.e. the post-merger and control have had the same cumulative feedback but the recent feedback history is different with mergers accreting more immediately post-coalescence

% \sout{By matching in black hole mass we prevent possible bias related to the AGN feedback model.}

\item The environmental parameter $N_2$ matched within a tolerance of $10\%$.

%\item log(SFR) within a tolerance of $0.01$ dex. 
 
%In Section~\ref{sec:bhmismatch}, we discuss how this requirement affects our results.
%As mentioned in section \ref{sec:AGNmodel}, above a threshold mass of M$_\text{BH} \sim 10^{8.2}$ M$_\odot$, the momentum injected into the medium surrounding the black hole by kinetic AGN feedback produces outflows of gas, that eventually will quench the galaxy. 
\end{itemize}

If more than one control is found for a given post-merger, we follow the weighting scheme of \citet{Patton2016} to select up to $5$ control galaxies. 
We then define the post-merger's \textit{control galaxy} as the single best control galaxy that shares the most number of subsequent snapshots with the post-merger's descendants whilst maintaining stellar and black hole mass match within a tolerance of $0.2$ dex of the descendants'. 
With our one-control-per-post-merger selection (fiducial matching scheme, in the following), we obtain robust control samples of $1440$, $1047$, and $548$ \eag{}, \ill{}, and \tng{} non-interacting galaxies, respectively. 
We note that we did not find any control galaxy for $111$, $35$, and $22$ post-mergers in \eag{}, \ill{}, and \tng{}, respectively. We exclude these post-mergers without control galaxies from the following analysis. 
We investigated the bias in the matched post-mergers of the three simulations. We found our matching scheme preferentially failed to find controls in post-mergers with higher stellar and black hole mass. Moreover, a few tens of \eag{} unmatched post-mergers have intermediate stellar mass with black hole masses that are either in the lower (in most of such cases) or higher envelope of the stellar mass vs black hole mass distribution. We stress that the matching criteria are necessary in order to select an unbiased set of non-interacting control galaxies, therefore, we prioritise the selection of unbiased control samples over the full completeness of the post-merger samples.
However, our choice of the maximum tolerances required in our matching criteria returns robust samples of mergers with completeness around $\mathbf{93\%}$ in \eag{} and above $96\%$ in \ill{} and \tng{}. 
Finally, we note that in Section~\ref{sec:observations}, we will analyse two matching scheme variations that include a complete sample of post-mergers that are not matched in black hole mass.

%In the $\sim 97\%$ of cases we found at least one control galaxy for every star-forming PM. In the other cases, we increase the tolerance range (and, if required, up to a maximum of) by 0.05 (0.1) dex for the M$_\ast$, 0.01 (0.04) dex for the SFR, 0.025 (0.1) dex for the M$_\text{BH}$ and 10\% (20\%) for both $N_2$ and $r_1$. 
%If more than one control is found, we follow the weighting scheme of Patton et al. (2016) to select up to five best matches in all the parameters. 
%In these cases, we choose the control galaxy to be the one among the best five that has the maximum number of subsequent snapshots in common with the post-merger, and that keeps a match with the post-mergers descendants in the environmental parameters $N_2$ and $r_1$, within a tolerance of 40\%. 
%Figure~\ref{fig: shows the distributions of redshift, M$_\ast$, $N_2$,  $r_1$, SFR, and  M$_\text{BH}$ for the star-forming post-mergers and their control galaxies. 
%$The matching process offers a control population that %well matches our post-merger sample in all the aforementioned parameters.

The three insets in the right panels of Figure~\ref{fig:distributions} show the distribution of the differences (i.e. post-mergers values $-$ control galaxy values) in the three matched parameters (note that redshift is matched de facto as well), demonstrating that the differences between the two samples of mergers and controls are well confined within the tolerance limits.

Once the control galaxy sample has been identified, we follow the evolution of the SFR in descendants of  %with the time elapsed from the coalescence of the $\Delta$SFR of both post-mergers descendants and their controls.
post-mergers and controls forward in time through the simulation. 
For each galaxy, we evaluate the relative level of star formation by calculating, on a logarithmic scale in the SFR-M$_\ast$ plane, the vertical offset between its SFR and the SFMS (hereafter, $\Delta$SFR).
Then, we define galaxies to be quenched when their $\Delta$SFR drops below $-0.9$ dex (i.e., a deviation of $<-3 \sigma$ from the SFMS). Additionally, we quantify the enhancement or deficit of AGN activity in post-mergers as the logarithmic difference between the SMBH accretion rate of each post-merger galaxy and the best matched control (hereafter, $\Delta \dot M_{BH}$), where a positive $\Delta \dot M_{BH}$ indicates a post-merger SMBH accretion rate $10^{\Delta \dot M_{BH}}$ higher than the control.
%- Reminds of TNG300 results \citepalias{Quai2021}.
%- Comparison Illustris, TNG, EAGLE.

%%%%%%%%%%%%%%%%%%%%%%%%%%%%%%%%%%%%%%%%%%%%%
%%%%%%%%%%%%%%%%%%%%%%%%%%%%%%%%%%%%%%%%%%%%%
%%%%%%%%%%%%%%%%%%%%%%%%%%%%%%%%%%%%%%%%%%%%%
%\section{BH accretion enhancement}
%- Reminds of TNG100 results (from Shoshannah)

%- Comparison Illustris, TNG, EAGLE

%%%%%%%%%%%%%%%%%%%%%%%%%%%%%%%%%%%%%%%%%%%%%
%%%%%%%%%%%%%%%%%%%%%%%%%%%%%%%%%%%%%%%%%%%%%
%%%%%%%%%%%%%%%%%%%%%%%%%%%%%%%%%%%%%%%%%%%%%
\section{Results}
\label{sec:results}
%%%%%%%%%%%%%%%%%%%%%%%%%%%%%%%%%%%%%%%%%%%%%
%%%%%%%%%%%%%%%%%%%%%%%%%%%%%%%%%%%%%%%%%%%%%
%%%%%%%%%%%%%%%%%%%%%%%%%%%%%%%%%%%%%%%%%%%%%

%%%%%%%%%%%%%%%%%%%%%%%%%%%%%%%%%%%%%%%%%%%%%
%%%%%%%%%%%%%%%%%%%%%%%%%%%%%%%%%%%%%%%%%%%%%
\subsection{Merger-driven AGN activity}
\label{sec:accretion}

% \textcolor{red}{Shoshanna's text and figure here.}
%%%%%%%%%%%%%%%%%%%%%%%%%%%%%%%%%%%%%%%%%%%%%
%%%%%%%%%%%%%%%%%%%%%%%%%%%%%%%%%%%%%%%%%%%%%
% \sbcom{Stats for ref when writing section}
% \begin{itemize}
%     \item EAGLE, median enhancement 0.39, positive enhancement 794/1216 or 65percent
%     \item Illustris, median enhancement 0.63, positive enhancement 525/668 or 78percent
%     \item TNG, median enhancement 0.33, positive enhancement 334/410 or 81percent 
% \end{itemize}

\begin{figure}
\centering
\includegraphics[width=1\columnwidth]{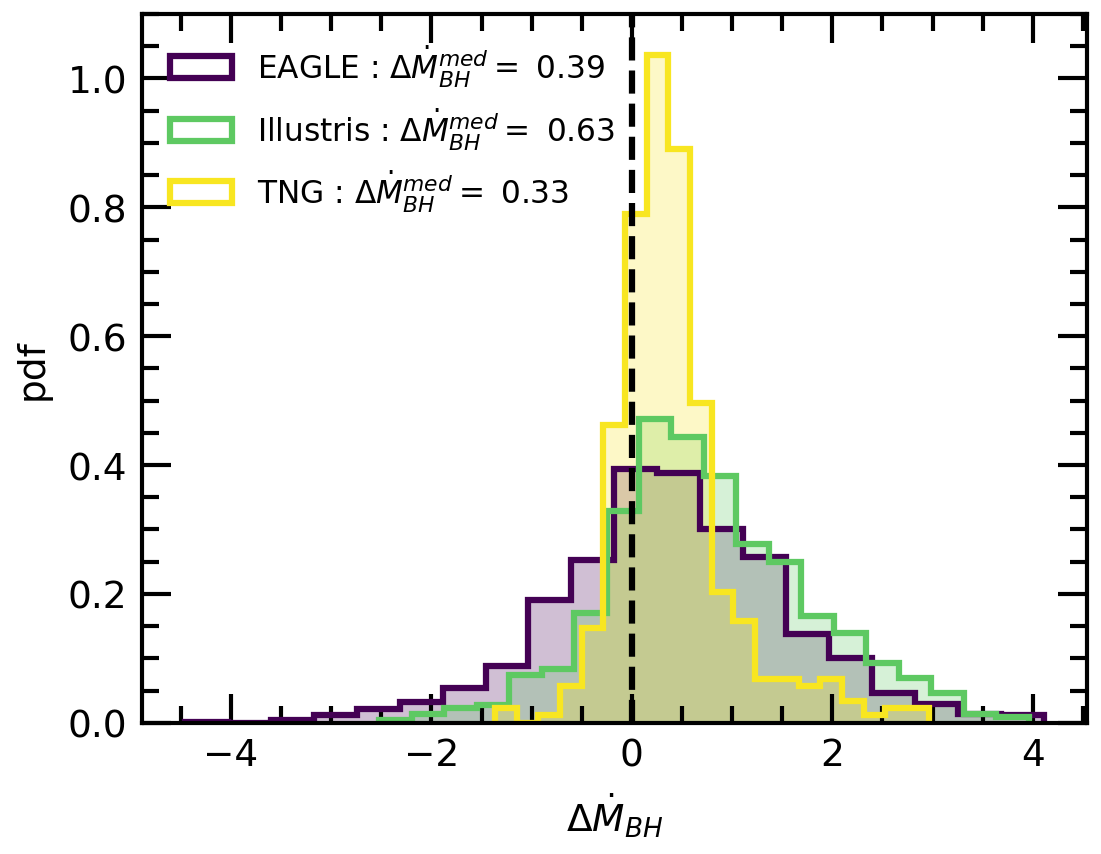}
\caption{Histogram of the relative offset between the SMBH accretion rates of star forming post-mergers and their best matched controls ($\Delta \dot M_{BH}$). Colours represent different simulations, purple for \eag{}, green for \ill{}, and yellow for \tng{}.
The figure demonstrates that, on average, the post-mergers from all the three simulations have higher accretion rates then their matched controls. The figure also demonstrates that a significant number (ranging from 20-30\%) of post-mergers have similar or lower SMBH accretion rates than controls.}
\label{fig:SMBHARenhancements}
\end{figure}

The primary goal of this work is to investigate the interconnection between galaxy mergers, AGN activity and star formation quenching in three state-of-the-art cosmological simulations with publicly available data, i.e., \eag{}, \ill{} and \tng{}. We begin by investigating the enhancement of AGN activity in the star forming post-merger population. Figure \ref{fig:SMBHARenhancements} shows the distribution of the logarithmic difference ($\Delta \dot M_{BH}$) between the SMBH accretion rates of post-mergers and controls for each of the three simulations. For the \eag{} simulation, shown in purple, post-mergers have a median accretion rate 2.5 times higher than controls. For the \ill{} simulations, shown in green, post-mergers have a median accretion rate 4.3 times higher than controls. Finally, for the \tng{} simulation, shown in yellow, post-mergers have a median accretion rate 2.1 times higher than controls, consistent with the results of \cite{ByrneMamahit2022} who find post-mergers undergoing quasar mode feedback (the majority of which are star forming) have accretion rates 1.9 times higher than controls. Therefore, Figure \ref{fig:SMBHARenhancements} demonstrates that post-mergers have, on average, higher accretion rates than controls in all three simulations. We find that \eag{} and \tng{} have quantitatively similar enhancements in post-mergers. In contrast, \ill{} post-mergers have significantly stronger enhancements of accretion rate. The stronger enhancement in \ill{} is consistent with the overabundance of efficiently accreting SMBHs in \ill{} \citep[][]{Habouzit2022}.  In addition to the enhancement in post-mergers, our results demonstrate that enhanced SMBH accretion rates (when compared with controls) are not ubiquitous, and a significant portion (ranging from 20 to 35 percent) of post-mergers have similar or lower accretion rates compared with their matched controls. We therefore demonstrate that the majority, but not all, of the star forming post-merger population experiences accretion rates in excess of their matched controls. 

%Figure \ref{fig:SMBHARenhancements} additionally highlights the incidence of elevated accretion rates in the star forming post-mergers, where approximately 65\% of \eag{}, 78\% of \ill{}, and 81\% of \tng{} post-mergers have an SMBH accretion rate higher than their controls. removed following Sara's comment

\begin{figure}
\centering
\includegraphics[width=1\columnwidth]{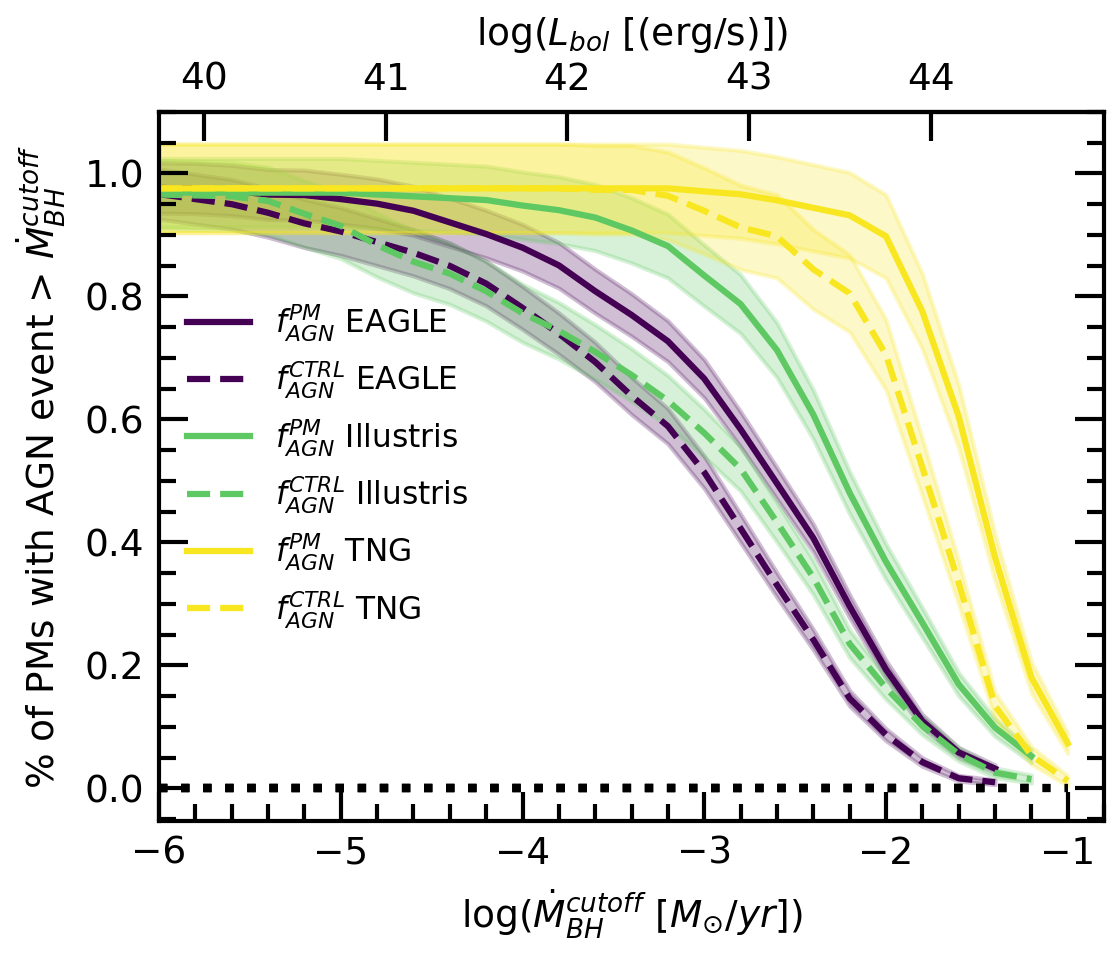}
\caption{Percentage of star forming post-mergers (solid lines) and controls (dashed lines) which experience an SMBH accretion rate event in excess of $\dot M_{BH}^{\text{cutoff}}$ within 500 Myrs post-coalescence. The top axis shows the equivalent bolometric luminosity, calculated as $0.1\dot M_{BH}c^2$. Colours represent different simulations, purple for \eag{}, green for \ill{}, and yellow for \tng{}.
The figure demonstrates that instances of high SMBH accretion rates occur more frequently in the post-mergers compared with the controls. }
\label{fig:fracAGNEvent}
\end{figure}

\begin{figure}
\centering
\includegraphics[width=1\columnwidth]{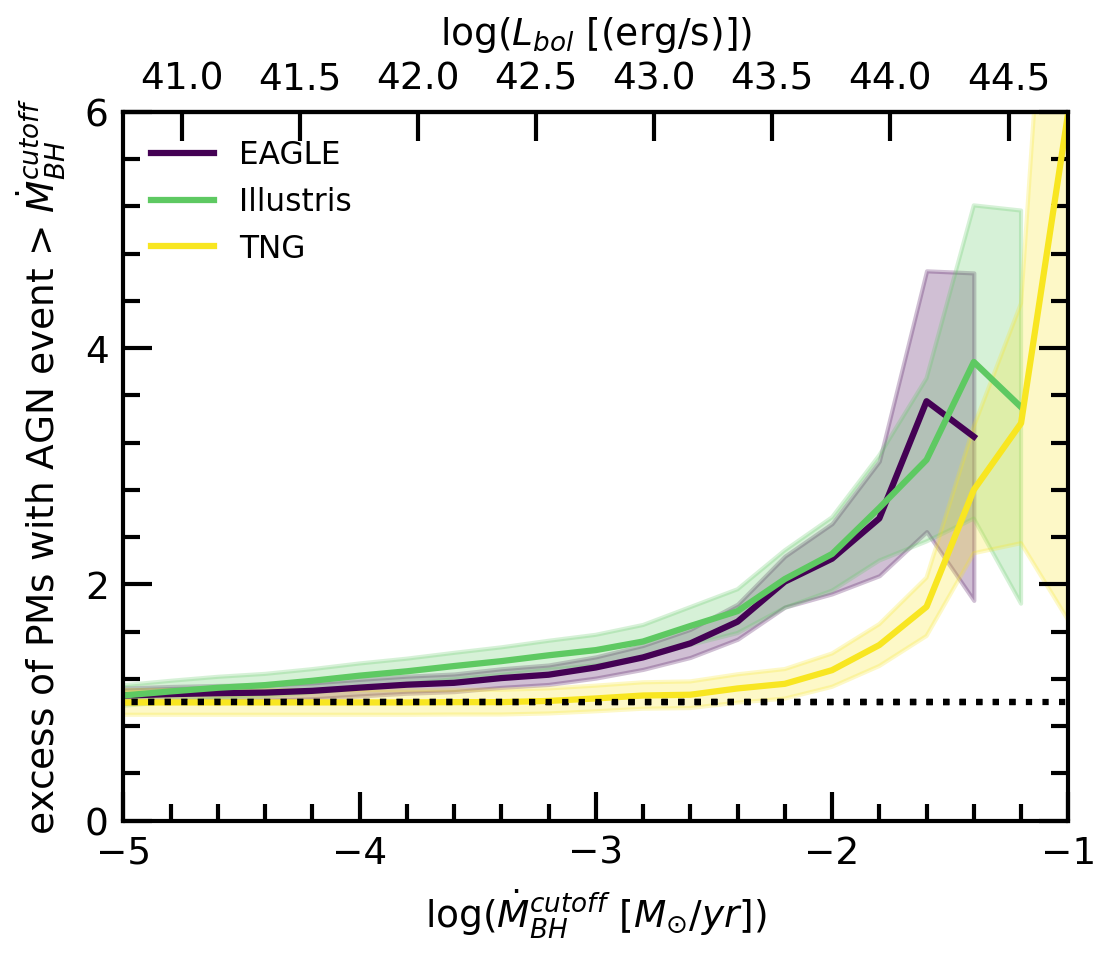}
\caption{Fractional excess of SMBH accretion rate events (within 500 Myrs post-coalescence) in the post-mergers relative to the controls. The top axis shows the equivalent bolometric luminosity, calculated as $0.1\dot M_{BH}c^2$. Colours represent different simulations, purple for \eag{}, green for \ill{}, and yellow for \tng{}. The excess is only calculated for values of $\dot M_{BH}^{\text{cutoff}}$ where there are more than 20 post-merger and non-merger galaxies, to ensure statistical robustness. The figure demonstrates that there is an excess of up to 3-4 times more highly accreting SMBHs in the post-merger population.}
\label{fig:fracAGNExcess}
\end{figure}

We further investigate the incidence of high accretion rate events within the star-forming post-merger population with the following procedure introduced in \citet{ByrneMamahit2022}. For each post-merger galaxy and its associated control, we follow the post-merger and control forward in time 500 Myrs, for intervals of time following the snapshot resolution of the respective simulations. We then determine if, for any snapshot within 500 Myrs, the post-merger and/or control has an SMBH accretion rate in excess of a given AGN threshold accretion rate ($\dot M_{BH}^{\mathrm{cutoff}}$). We note that we only conduct this experiment for post-mergers which can be followed forward in time for 500 Myrs. Repeating the experiment for each post-merger in the sample yields the percentage of post-mergers (or controls) that have undergone a high accretion rate event within 500 Myrs. Figure \ref{fig:fracAGNEvent} demonstrates the results for \eag{}, \ill{}, and \tng{}, where post-mergers are shown as solid lines and controls as dashed lines. The y-axis demonstrates the percentage of post-mergers or controls which undergo an `AGN event', and the x-axis defines the minimum SMBH accretion rate cutoff which defines the `AGN event'. Figure \ref{fig:fracAGNEvent} demonstrates that AGN events become more rare with increasing AGN cutoff luminosity. We find that very few ($\sim 6\%$) of \eag{} post-mergers have (snapshot averaged) accretion rates above $L_{bol} \sim 10^{44}$ erg/s. Similarly, 17\% of \ill{} post-mergers have accretion rates above $L_{bol} \sim 10^{44}$ erg/s. In TNG, 60\% of post-mergers have $L_{bol} > 10^{44}$ erg/s in \tng{}, but there is a rapid decline with no significant amount achieveing $L_{bol}>10^{45}$ erg/s. 
We note that by design, the simulations limit high accretion rate events, both through the implementation of Eddington limited accretion rates and the transition to effective AGN feedback in TNG at a black hole mass of approximately $10^{8.2}$ M$_\odot$.
%\textbf{although the effect may be attributable to the built-in switch mechanism to low accretion mode at black hole mass of M$_\text{BH} \geq 10^{8.2}$ M$_\odot$.} 
We therefore find highly accreting SMBHs are rare in \eag{} and \ill{}. However, despite the rarity of the events, they occur more frequently in the post-merger population compared with the control population. We note that the average time between snapshots is shortest for \eag{}, which could result in higher average accretion rates, however we find instead that \eag{} produces the fewest galaxies with high SMBH accretion rates. We additionally note that although high accretion rate events are rare in post-mergers, it does not mean that the merging galaxy has not undergone any period of rapid accretion, which may still occur in the pre-merger or ongoing merger phase. In Figure \ref{fig:fracAGNExcess} we show the fractional excess (the ratio of the solid to dashed lines in Figure \ref{fig:fracAGNEvent}) as a function of AGN accretion rate. In all three simulations, post-mergers more frequently host highly accreting SMBHs, with an increasing excess as a function of SMBH accretion rate. \eag{} and \ill{} demonstrate a peak excess of approximately four, that is four times more post-mergers achieve SMBH accretion rates of $ L_{bol}> 10^{44}$ erg/s compared with controls. Similarly, post-mergers from \tng{} achieve a peak excess of approximately five to six, although at higher uncertainty. The fractional excess in \tng{} we find here is higher than the excess from \citet{ByrneMamahit2022}, but is consistent with a stronger enhancement of SMBH accretion rate in the star-forming post-merger sample. Our results are also comparable with \citet{McAlpine2020}, who find a maximum AGN excess (comparing post-mergers to a control sample) of approximately two in \eag{}. In fact, it is expected we may find a slightly higher excess over \citet{McAlpine2020}, since we are accounting for any AGN event occuring in the snapshots between coalescence and 500 Myrs post-coalescence. Overall, Figures \ref{fig:fracAGNEvent} and \ref{fig:fracAGNExcess} demonstrate in combination that highly accreting SMBH events are rare in the star forming post-merger population but occur more frequently than in the control population.

%%%%%%%%%%%%%%%%%%%%%%%%%%%%%%%%%%%%%%%%%%%%%
%%%%%%%%%%%%%%%%%%%%%%%%%%%%%%%%%%%%%%%%%%%%%
\subsection{Quenching in post-merger galaxies}
\label{sec:fractions} 
%%%%%%%%%%%%%%%%%%%%%%%%%%%%%%%%%%%%%%%%%%%%%
%%%%%%%%%%%%%%%%%%%%%%%%%%%%%%%%%%%%%%%%%%%%%
\begin{figure}
\centering
\includegraphics[width=1\columnwidth]{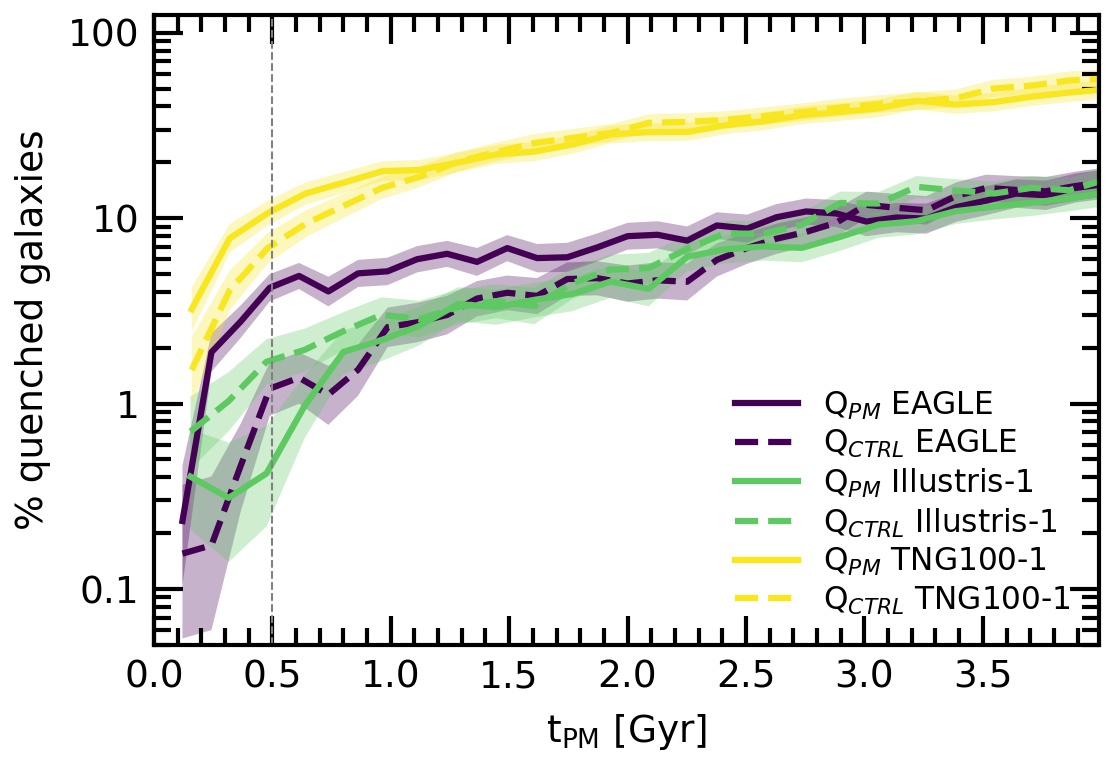}
\caption{Fraction (expressed in percentage) of post-merger galaxies (\QPM{}, solid curves)  and control galaxies (\QCTRL{}, dashed curves) that are quenching as a function of time after coalescence (\tpm{}).
Colours represent different simulations, purple for \eag{}, green for \ill{}, and yellow for \tng{}. 
The shaded regions represent the $1\sigma$ uncertainty on the fraction, estimated following \protect\citet[][]{Gehrels1986}.
We chose to represent the quenching fraction with a log y-axis scale, to highlight the trend shortly after the merger. 
The figure demonstrates that all the three simulations have a small fraction of post-merger quenching shortly after coalescence. }
\label{fig:fractions}
\end{figure}

Now that we have established the frequency of enhanced SMBH accretion in the post-merger samples, we continue our analysis by studying the impact (or lack thereof) of mergers on star formation quenching. Following \citet{Quai2021}, we quantify the effects of mergers using two metrics: (i) the fraction of post-merger descendants with quenched star formation, and (ii) the excess of quenched post-mergers, i.e., the number of quenched post-mergers normalized by the number of quenched controls as a function of time after coalescence. 
Combined, these two metrics allow us to quantify the absolute rate of quenching in post-mergers, as well as addressing whether quenching is more common in post-mergers than in controls.  

In Figure~\ref{fig:fractions}, we show the evolution of the fraction of post-merger galaxies (solid lines) that are quenched as a function of time elapsed since coalescence (t$_\text{PM}$). For a qualitative comparison, we show also the behaviour of the control samples (dashed line), that represent the secular incidence of quenching that is expected.
The trends of the three simulations are represented with different colours, purple for \eag{}, green for \ill{}, and yellow for \tng{}, with the shaded regions enclosing the $1\sigma$ uncertainty of the fraction, that we estimate following \citet[][]{Gehrels1986}.
We note that at t$_\text{PM} = 0$, the quenched fraction is $0$ because, by construction, we select star-forming post-mergers in our initial sample, hence we begin the analysis from the first snapshot after coalescence.

The quenched fraction of \eag{}’s post-mergers grows faster than their control galaxies in the vicinity of coalescence, though showing only a modest fraction of $\sim4.5$ percent around $500$ Myr after coalescence. Later, the increase slows down and levels off to the control quenched fraction (i.e. the expected quenching because of secular evolution) around $3$ Gyr after the merger. 
Looking at the trend from a complementary perspective, the result means that around $90$ percent of \eag{}'s post-mergers that were star-forming during coalescence keep forming stars long after the merger is complete.  

\ill{} shows very little impact of mergers on quenching star formation, as only $0.4$ percent of them quench within $500$ Myr after coalescence (against $\sim1.9$ percent of their control sample in the same period), when mergers are supposed to manifest the maximum effect on the evolution of star formation \citep[e.g.,][]{Hani2019}. 
However, the \ill{} trend shows a faster growth in the rate of quenching than \eag{}, thus catching up with the control sample trend around $1$ Gyr after the merger, and reaching a quenched fraction $\sim10$ percent $4$ Gyr after the merger. 

In \tng{}, we find the largest quenched fraction compared to the other simulations, although reaching only around $10$ percent within $500$ Myr after the merger, and levelling off with the control sample trend shortly after one Gyr after coalescence. 
The difference in the quenched fraction trend between \tng{} and the other two simulations can be better understood by focusing only on the control sample behaviour. We find that \eag{} and \ill{} control samples show a quantitatively similar trend for the entire period analysed, whilst the trend of the \tng{} control sample is systematically above, thus suggesting that the main distinction can be ascribed to both a different star formation/quenching regulation in the simulation models \citep[e.g.,][]{Donnari2019} and to our selection of \tng{} galaxies that already have massive black holes about to enter (or already have entered) into the effective kinetic mode regime \citep[i.e., M$_\text{BH}>10^{8.2}$M$_\odot$, see Figure~\ref{fig:distributions}, and][]{Terrazas2020}. Hence the controls quench just as readily as the merging galaxies.

\begin{figure}
\centering
\includegraphics[width=1\columnwidth]{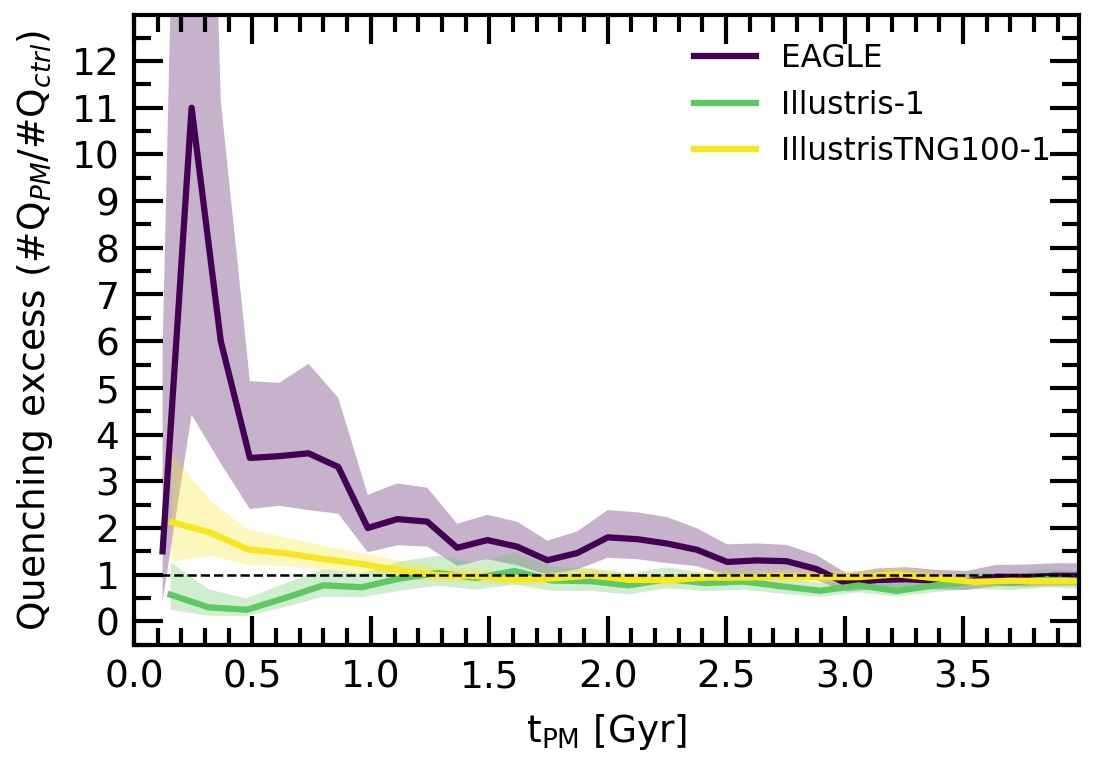}
\caption{Excess of quenched post-merger galaxies as a function of time after coalescence. The excess is defined as the ratio between the number of quenched post-mergers ($\#$\QPM{}) and quenched control galaxies ($\#$\QCTRL{}).
Colours represent different simulations, purple for \eag{}, green for \ill{}, and yellow for \tng{}. 
The shaded contours represent the $1\sigma$ uncertainty, that is quantified following \protect\citet[][]{Gehrels1986}. 
The horizontal dashed black line represents the same number of quenched post-mergers and controls. The figure indicates that \eag{} shows the largest excess of quenched post-mergers amongst those analysed in this work, with a factor of around $11$ times more quenched galaxies than expected from the non-interacting population. Conversely, \ill{} shows a deficit of quenching in post-merger galaxies. Finally, \tng{} shows an excess of quenched post-mergers, qualitatively and quantitatively comparable with that in \protect\citetalias{Quai2021}.
}
\label{fig:excess}
\end{figure}
Figure~\ref{fig:excess} shows the quenching excess, that quantifies the relative tendency of mergers to experience quenching compared to secular processes which are accounted for in the controls.
The layout is the same as in Figure~\ref{fig:fractions}, and the uncertainties are from the binomial statistic \citep[][]{Gehrels1986}.
We find that \eag{} (represented in purple)  shows the largest excess among the analysed simulations, with a maximum of $11$ times more quenched post-mergers than expected in a control sample around $250$ Myr after coalescence. 
Then, the excess decreases steadily until $\sim3$ Gyr after the merger, when quenching in post-mergers becomes statistically indistinguishable from that of the control sample. 
%\tng{} (represented in yellow), shows an %excess, as well as \eag{},  though %finding that quenching occurs in \tng{} %post-mergers at twice the rate of the %control galaxies. 
\tng{} (shown in yellow) also shows an enhanced quenched fraction, but at a somewhat lower level (factor of two) than in \eag{}.
Moreover, the excess vanishes more quickly than in \eag{}, with the number of quenched post-mergers compatible with that expected from secular evolution beyond $1.2$ Gyr after coalescence.  \ill{} (represented in green), surprisingly shows a deficit, with quenching happening in post-merger galaxies at half the rate expected in controls. The presence of a deficit in quenched post-mergers means that in \ill{} mergers are actually halting, or slowing down, the quenching process. 
The effect of reducing the frequency of quenching is temporary, and the deficit disappears around $1$ Gyr after the merger, when the number of post-mergers and controls is statistically equal. 
\subsection{Comparing post-merger quenched fractions with observations}
\label{sec:observations}

The picture that arises from our analysis in the previous sections is that mergers in state-of-the-art cosmological simulations may trigger AGN, however, quenching occurs in only a small fraction of merging systems.
Although we find a small fraction of promptly quenched galaxies (Figure~\ref{fig:fractions}), nonetheless there is an excess of quenched post-mergers compared with control galaxies in \eag{} and \tng{} (Figure~\ref{fig:excess}), thus suggesting that mergers in these simulations could contribute to quenching star formation.
We stress that our experiment proves how contemporary subgrid models affect the evolution of quenching in simulated merger remnants \citep[see Section~\ref{sec:data}, and][]{Quai2021}. 
However, to obtain more accurate and reliable conclusions that can be generalised to the actual Universe, we need to put our results into perspective with robust observational results.

In the Introduction, we mentioned that `post-starburst' galaxies are a useful population for identifying rapid and recent quenching. 
Briefly, post-starbursts are commonly identified by their optical spectra dominated by an over-abundance of A and F types stars (with a lifetime around $1$ Gyr), and a lack of spectral features of shorter-lived massive 0 and B types stars, indicating a recent
starburst episode followed by a sharp truncation of star formation \citep[e.g.,][]{Dressler1983, Couch1987}. Post-starbursts could therefore represent the optimal population to causally connect mergers and quenching. 

\begin{figure}
\centering
\includegraphics[width=1\columnwidth]{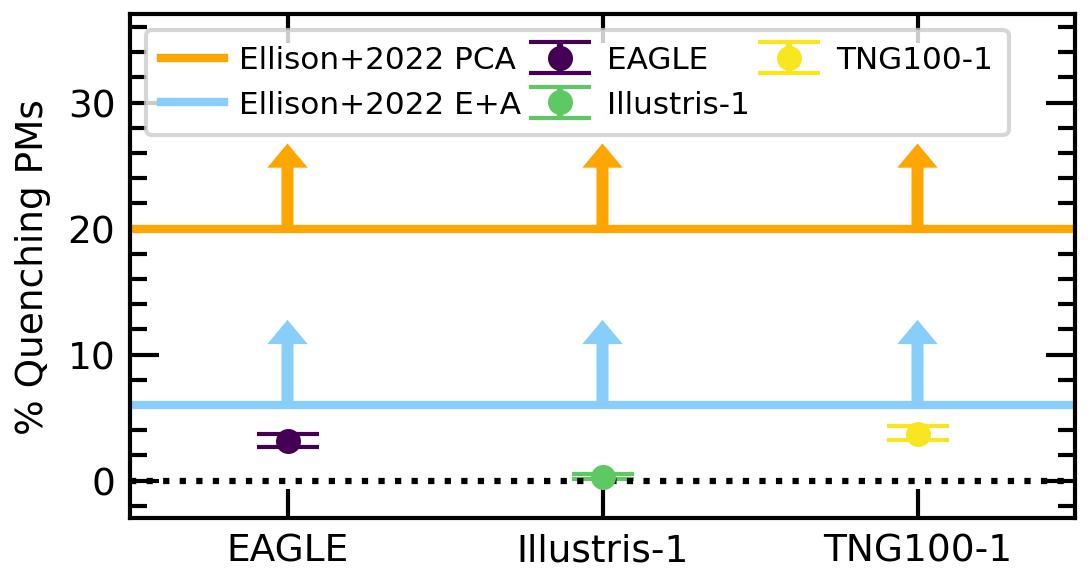}
\includegraphics[width=1\columnwidth]{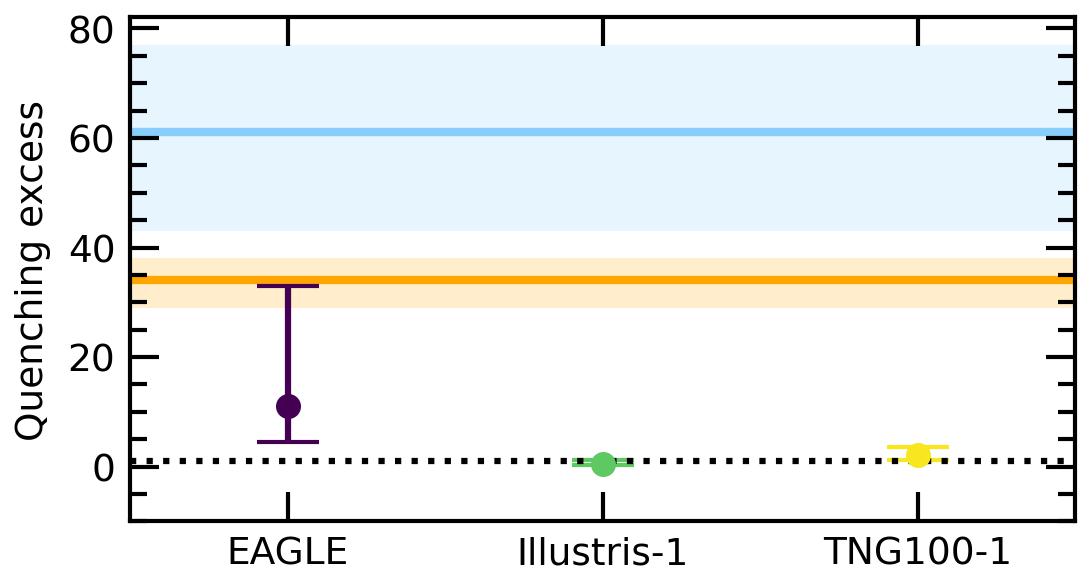}
\caption{Top: comparison of the global quenched fraction in simulated post-mergers with the post-starburst fraction in mergers from observations in \protect\cite{Ellison2022} (the arrows indicate that values are lower limits, the light blue line represents observed post-starbursts in mergers selected using a classical E+A method, whereas the orange line is for post-starbursts in mergers selected applying a PCA approach). 
The three symbols and errorbars represent the quenched fraction within $500$ Myr from coalescence of, from left to right, \eag{} (purple), \ill{} (green), and \tng{} (yellow). 
Bottom: comparison of the quenching excess in simulations with observations. The layout is the same as in the top panel. 
The figure demonstrates that cosmological simulations like the ones analysed in this paper cannot quantitatively reproduce observational constraints of the impact of mergers on quenching star formation.
}
\label{fig:comparis}
\end{figure}

In the literature, however, most studies focus on the fraction of mergers in the post-starburst population, finding values ranging from $10$ to $60\%$ \citep[e.g.,][]{Blake2004, Pracy2009, Setton2022, Wilkinson2022}. \citet{Wilkinson2022} showed that the post-starburst merger fraction is strongly susceptible to the choice of merger identification metrics and post-starburst selection criteria. \citet{Wilkinson2022} also demonstrated that only $\sim30\%$ of recent mergers in simulations can be recovered using the typical morphology metrics.
Therefore, it cannot be excluded that a larger fraction of
post-starbursts could have
recently undergone a merger. 
Nevertheless, even in an extreme scenario where nearly all post-starburst galaxies would be post-mergers, it would not answer the complementary
question of whether mergers systematically lead to promptly quenched star formation.

Recently, \citet{Ellison2022} analysed a robust and highly pure sample of $\sim 700$ post-mergers identified by \citet{Bickley2022}, the largest pure catalogue to date, allowing them to statistically tackle the frequency of post-starburst galaxies in recent mergers.
Applying two different selection methods for detecting post-starbursts, i.e., the `E+A' and the `Principal Component Analysis' (PCA)\footnote{The PCA selects galaxies with a relative excess of Balmer absorption lines, thus indicating spectra dominated by A and F types stars, but the method includes also galaxies that show emission lines that might originate from a non-star-forming source \citep{wild2007}, which would be instead removed in the E+A selection \citep[e.g.,][]{Zabludoff1996, Goto2005}.}, \citet{Ellison2022} found, respectively, a frequency of $6$ and $20$ percent of post-starbursts in the post-merger population. 
Compared with a control sample of non-mergers, these post-starburst fractions represent a factor $30-60$ enhancement, demonstrating that mergers can lead to quenching.
%Therefore, the E+A post-starburst sample is highly pure %but incomplete, whilst the PCA post-starburst sample is %highly complete but it could also be contaminated by %non-genuine post-starburst galaxies, hence, the %observational quenching fraction from the PCA approach %could be either a lower or an upper limit to the actual %value.  \sbcom{I'll hold off on comments on this since %you may make changes after Sara's comments at the %meeting}}, \citet{Ellison2022} found, respectively, a frequency of $6$ and $20$ percent of post-starbursts in the post-merger population. 

However, \citet{Ellison2022} stress that these observational fractions of post-starbursts in post-mergers are lower limits of the actual quenching fraction for the following reasons.
First of all, to pass through the post-starburst phase, a galaxy has to first experience an enhanced star formation phase. However, we know from observations that the majority of post-mergers in the local Universe show a relatively modest enhancement in star formation \citep[e.g.,][]{Ellison2013, Knapen2015}.
Therefore, many post-mergers could quench as a direct consequence of the interaction without passing through the post-starburst phase.
A second critical point concerns the different timescales at play. 
Galaxy merger features (e.g., morphology disturbance, shells, tidal features, star formation enhancement) tend to vanish on the order of $200-500$ Myr after coalescence \citep[e.g.,][]{Mihos1995, Lotz2008, Hani2019, Bottrell2022}.
The post-starburst phase, instead, arises typically $0.5-1$ Gyr after the burst of star formation \citep{Wild2020}.
Therefore, many post-mergers might be prematurely observed before the onset of the post-starburst phase. The chance to observe a post-starburst phase in post-merger galaxies scales with a restricted combination of orbital properties, mass ratio and underlying properties of the galaxies themselves. 
Mergers with prograde orbits, high mass ratio, and high initial gas mass are able to induce stronger and longer-lasting morphology disturbance and quenching \citep[e.g.,][]{DiMatteo2007, Lotz2008, Moreno2015, Bekki2005, Wild2009, Pawlik2018,
Zheng2020}.
Indeed, \citet{Ellison2022} note an enhanced post-starburst fraction in more asymmetric mergers.

%Finally, the post-starburst fraction of a merger sample %is sensitive to both merger identification methods and %post-starburst sample selection criteria \citep[see][, %for a thorough investigation of the impact of different %metrics on merger identification and post-starburst %selection]{Wilkinson2022}. \sbcom{I think you could put %this in the paragraph where you comment on the same %paper and the effect on the merger fraction of a PSB %sample in order to reduce repetition}

%The PCA selects galaxies with a relative excess of %Balmer absorption lines, thus indicating spectra %dominated by A and F types stars, but the method %includes also galaxies that show emission lines that %might originate from a non-star-forming source %\citep{wild2007}, which would be instead removed in %the E+A selection \citep[e.g.,][]{Zabludoff1996, %Goto2005}. 
%Therefore, the E+A post-starburst sample is highly pure %but incomplete, whilst the PCA post-starburst sample is %highly complete but it could also be contaminated by %non-genuine post-starburst galaxies, hence, the %observational quenching fraction from the PCA approach %could be either a lower or an upper limit to the actual %value.  \sbcom{I'll hold off on comments on this since %you may make changes after Sara's comments at the %meeting}

In Section~\ref{sec:fractions}, we measured the fraction of quenched galaxies in the post-merger populations of the three analysed cosmological simulations. We cannot however directly compare our quenched fractions to those of \citet{Ellison2022} because we define the fraction to be the ratio between the number of quenched post-mergers over the number of total post-mergers selected, which by construction belong to the sub-group of star-forming post-mergers at coalescence. 

In order to compare with \citet{Ellison2022}, we now quote the global quenched fraction, i.e. the number of star-forming post-mergers that quench relative to the total sample of post-mergers which include both the star-forming and non-star-forming populations.
%\sbcom{You could maybe do a subsubsection on %alternative matching schemes to add structure to break %up the long text}
%Another problem that we need to address regards the %different matching scheme applied for selecting control %galaxies. \sbcom{Just my preference, but I don't like %'problem' haha. I think you could remove the first %sentence and change the start of the next sentence from %'In fact' to 'In addition'}
Figure~\ref{fig:comparis} shows the global quenched fraction (top panel) and the quenching excess (bottom panel) within $500$ Myr after coalescence of the samples obtained with our fiducial matching selection (see Section 2.4), in comparison to observational results from \citet{Ellison2022}.
Figure~\ref{fig:comparis} shows that the quenched fractions measured in the three analysed simulations are below the lower limits of the observational values (both E+A and PCA selection), and significantly below the observational quenching excess values. Only EAGLE's quenching excess is barely compatible (within errors) with the excess measured with the conservative E+A selection. However, none of the simulations show quenched fraction and quenching excess compatible with the results from the PCA selection.

\begin{figure*}
\setlength\lineskip{-2pt}
\centering
\includegraphics[width=0.33\linewidth]{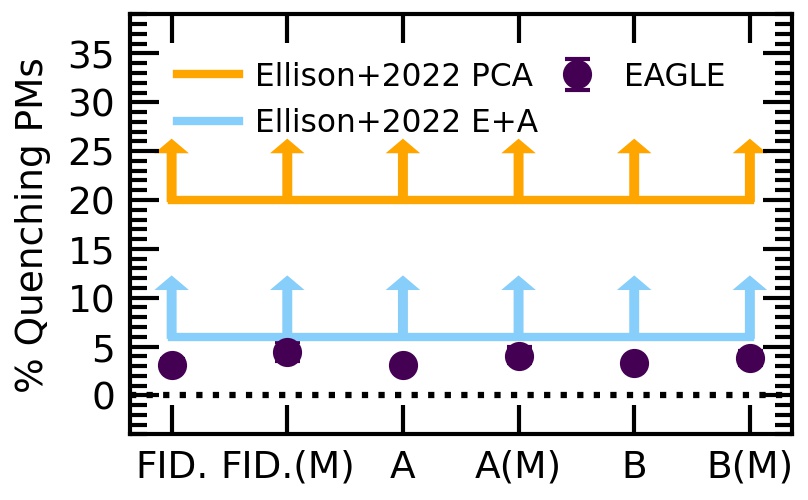}
\includegraphics[width=0.33\linewidth]{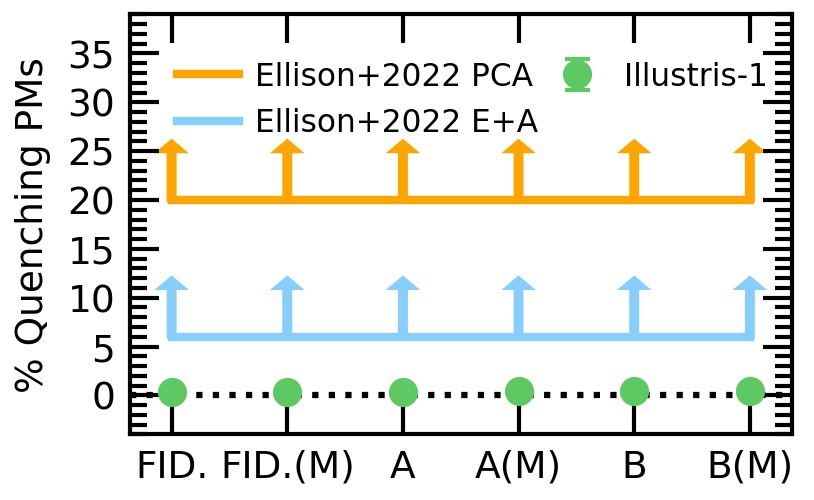}
\includegraphics[width=0.33\linewidth]{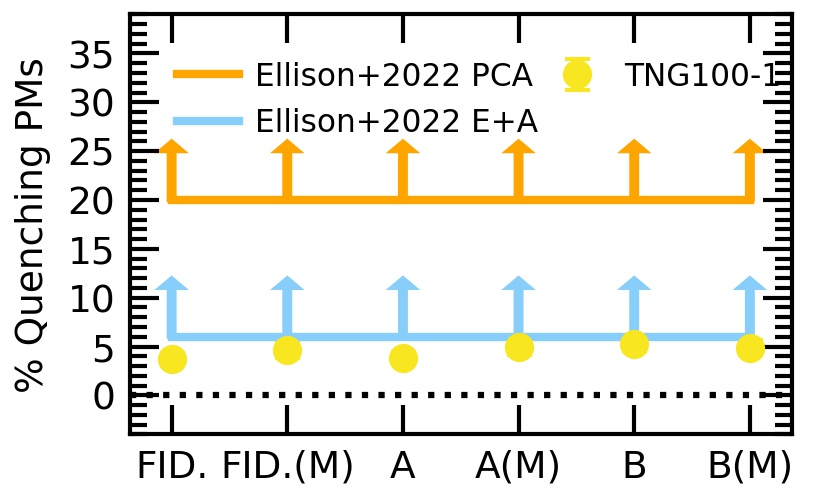}
\\
\includegraphics[width=0.33\linewidth]{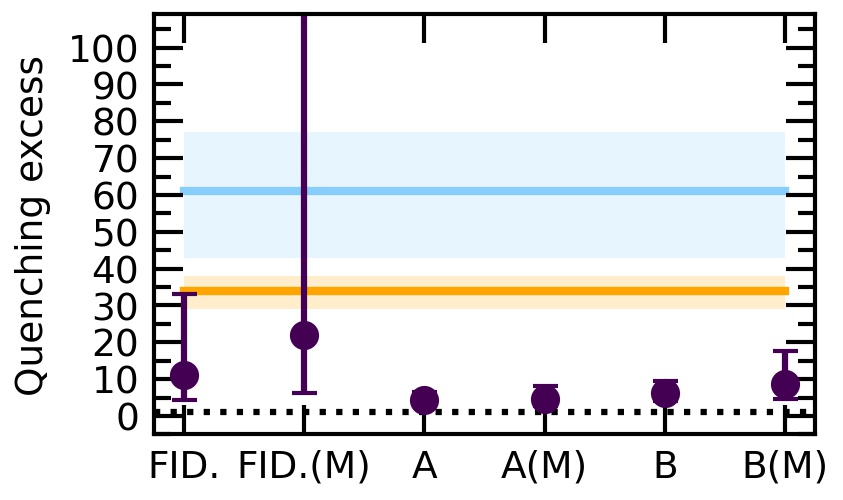}
\includegraphics[width=0.33\linewidth]{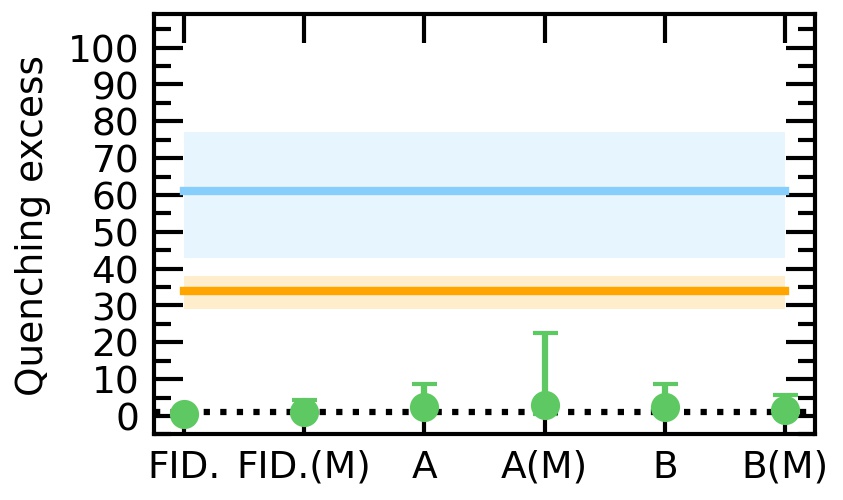}
\includegraphics[width=0.33\linewidth]{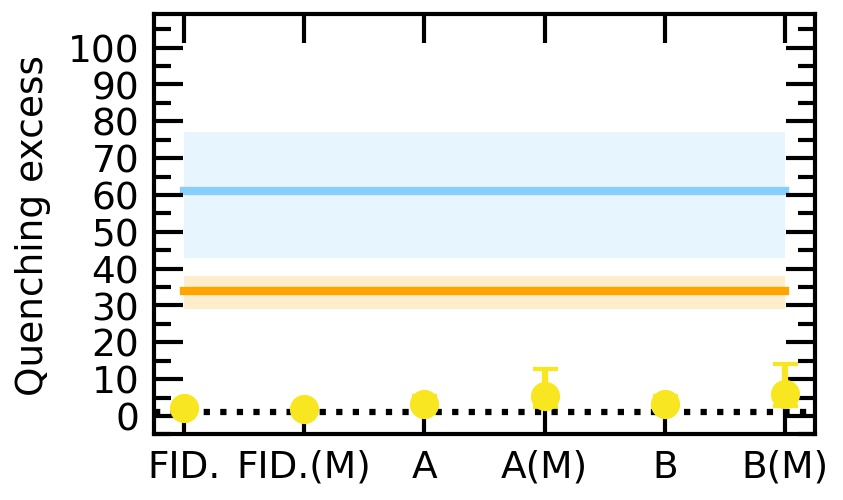}

\caption{Top: comparison of the global quenched fraction in simulated post-mergers with the post-starburst fraction in mergers from observations in \protect\cite{Ellison2022} (the arrows indicate that values are lower limits, the light blue line represents observed post-starbursts in mergers selected using a classical E+A method, whereas the orange line is for post-starbursts in mergers selected applying a PCA approach). 
Each panel is assigned to one of the three simulations, from left to right, \eag{} in purple, \ill{} in green, and \tng{} in yellow. 
Moreover, for each simulation, the comparison with observations is done using the outcome of six different matching schemes (see their definition in the text in Section~\protect\ref{sec:observations}).
Bottom: comparison of the quenching excess in simulations with observations. The layout is the same as in the top panels. 
The figure depicts how the analysed cosmological simulations cannot quantitatively reproduce observational constraints of the impact of mergers on quenching star formation.
}
\label{fig:observation}
\end{figure*}

We test whether the incompatibility between simulations and observational constraints is due to differences in the matching selection criteria.
Although the control sample is not used for measuring the quenched fraction, a different set of matching criteria can modify the completeness of the post-merger sample, because post-mergers that do not have a control are excluded from the analysis.
In \citet{Ellison2022} each post-merger is matched to the best
non-interacting control galaxy in redshift and stellar mass.
Here, we perform two variations of our fiducial matching scheme (see Section~\ref{sec:control_sample}), changing our criteria one step at a time to get closer to the observational control selection. 
For each variation, we take the maximum quenched fraction within $500$ Myr after coalescence, in order to focus on the stage of the merger sequence where it is highest the probability to observationally select galaxies with both post-merger and post-starburst features.
This experiment helps us to both compare the simulated and observed merger quenched fraction, and to measure the impact of the matching scheme on our results.
The fiducial and the two variation schemes are summarised in the following:
\begin{itemize}
\item \emph{Fiducial}. For each star-forming post-merger, we select the best star-forming non-interacting galaxy that is matched in redshift, stellar mass, black hole mass, and environment. See Section~\ref{sec:control_sample} for details.
\item \emph{Scheme A}. We exclude black hole mass, while we apply the matching criteria on redshift, stellar mass, and environment.
\item \emph{Scheme B}. We exclude black hole mass, and environment, while we keep the criteria on redshift and stellar mass. Scheme B offers the closest matching criteria to those in \citet{Ellison2022}, where each post-merger is matched to non interacting galaxies in redshift and stellar mass.
\end{itemize}
Additionally, \citet{Ellison2022} found that post-mergers showing the most disturbed morphologies exhibit a higher post-starburst fraction.
Because of the aforementioned reasons, it is likely that most observational post-starburst post-mergers might be remnants of major mergers.
Therefore, for each of the three matching variations, we also perform an additional experiment with a focus only on major mergers (i.e., mass ratio $> 1:4$).

The result of the experiment is presented in the top panels of Figure~\ref{fig:observation}. The three top panels show, from left to right, the merger quenched fractions measured in the various matching scheme variations for \eag{},\ill{},\tng{} respectively.
In each panel, the merger quenched fractions from \citet{Ellison2022} are reported in orange and light blue, for the PCA and E+A selections, respectively.
We also drew arrows to indicate that \citet{Ellison2022} values are lower limits for the reasons described earlier in this section.

\subsubsection{Global quenched fraction}
The top panels of Figure~\ref{fig:observation} present the global quenched fractions of the three simulations.  Symbols show the quenched fractions for the various matching schemes.  
Inspection of Figure~\ref{fig:observation} reveals that \eag{} post-mergers (top-left panel) show quenched fractions between $3.1^{+0.5}_{-0.4}$ and $4.4^{+1.1}_{-0.9}$ percent, with the larger values found in the variation scheme tests that are focused on major mergers.
\ill{} shows, instead, quenched fractions between $0.3^{+0.4}_{-0.2}$ and $0.5^{+0.3}_{-0.2}$ percent, values substantially smaller than observations in all the variation tests.
Similarly to \eag{}, the quenched fraction in \tng{} post-mergers (top-right panel) are between $3.7^{+0.6}_{-0.5}$ and $5.3^{+0.8}_{-0.7}$ percent.
We note that \tng{} quenched fraction changed significantly from that in Figure~\ref{fig:fractions}. Here, the \tng{} fraction goes down because of all the non-star-forming post-mergers introduced in the denominator.

The mild discrepancy between \eag{}, \tng{} and the E+A observational post-starburst fraction could be explained by a difference in quenching efficiency between simulations and observations. Previous works comparing simulations with observations show that the galaxy passive fraction is about $15\%$ too low in \eag{} \citep[][]{Furlong2015, Trayford2017}. Also in \ill{} quenching is not efficient enough compared to observations \citep[e.g.][]{Sparre2015, Bluck2016, Terrazas2017}, while the quenching mechanism is more efficient in \tng{} than previous simulations, hence showing a better agreement with observations \citep[e.g.,][]{Donnari2019}. 
The \eag{} and \tng{} quenched fractions seem to be quantitatively almost compatible with the lower limit of E+A selection (i.e., $\sim6\%$), with \tng{} showing a slightly better agreement than \eag{}. 
However, they are both substantially smaller than the observational PCA post-starburst fraction (i.e., $\sim20\%$), with a level of incompatibility that is arguably challenging to be explained by the different experimental set up.  

We finally note that the quenched fractions in the three analysed matching schemes (i.e., fiducial, A, and B) do not substantially differ from each other because the adopted criteria do not compromise the completeness of the post-merger sample, that is stably above $90\%$ in all the three simulations regardless of the matching approach. 
Finally, it is interesting to note that the variation tests focused on the major merger sub-samples show slightly larger quenched fractions than the corresponding tests on the whole merger sample, thus indicating that major mergers are more conducive to star formation quenching (see also Section~\ref{sec:analysis}). 

\subsubsection{Quenching excess}
The second part of our comparison with observations regards the comparison of the quenching excess (i.e. the ratio between the number of quenched post-mergers over the number of quenched control galaxies) between simulated and observed mergers. 
\citet{Ellison2022} found a strong statistical excess of post-starbursts in post-merger galaxies, with quenching being $34$ to $\sim60$ times (PCA and E+A selection, respectively) more common in post-mergers than in the control sample.
The bottom panels of Figure~\ref{fig:observation} show the quenching excess from our various matching scheme variations. The three panels refer, from left to right, to \eag{}, \ill{}, and \tng{}, respectively. In all the panels, we also report the observational excess of the PCA (in orange, with an excess of $\sim34^{+4}_{-5}$) and E+A (in light blue, with an excess of $\sim60^{+16}_{-18}$) selections from \citet{Ellison2022}.
None of the simulations has a quenching excess high enough to be comparable with the observations. 
\eag{} shows excess between $4.2^{+2.4}_{-1.5}$ and $22^{+110.6}_{-15.9}$ (in the scheme A and the fiducial scheme focused on major mergers, respectively). The latter, though compatible with observations within errors, have a very large uncertainty due to the paucity of quenched controls, i.e., only one quenched control galaxy versus $22$ quenched post-mergers in the same period of time.
\ill{} shows a post-merger quenching excess between $0.6^{+0.7}_{-0.3}$ and $3.0^{+19.7}_{-2.4}$, with values, on average, compatible with a deficit of quenching (i.e. quenching excess $<1$), which means quenching in \ill{} post-mergers is less common than expected in a control sample, thus consistent with what we already found in Section~\ref{sec:fractions}. 
Finally, \tng{} shows a post-merger quenching excess between $2.0^{+0.9}_{-0.6}$ and $5.3^{+7.4}_{2.9}$ (in the fiducial matching scheme focused on major mergers and in the scheme A focused on major mergers), with a slightly worse agreement with observations than \eag{}. 
Earlier in this section, we demonstrated that the quenched fractions in \eag{} and \tng{} post-mergers are quantitatively almost compatible with the E+A selection, hence, the incommensurable difference in the quenching excess is driven by a larger number of control galaxies that quench in the simulations than in the observations. 
The difference in the simulated and observational controls could be explained if most of the simulated controls quench because of secular reasons and only a few pass through a post-starburst phase. 
%\sbcom{This could, in principle, apply to PMs as well right?} 
However, we cannot test this speculative scenario because it would be challenging to isolate post-starburst galaxies in our global sample of mergers. 
Selecting a simulated sample of bona fide post-starbursts would require a different experimental approach and some fine-tuning to translate accurately the spectral post-starburst features into the star formation histories of the simulated galaxies (e.g., what should be the minimum strength of a burst to lead a galaxy through a post-starburst phase?). 

% \sbcom{I feel like this paragraph ends on a low point (i.e. you say maybe the excesses don't match because we need to better select PSBs in the simulation instead of just rapid quenchers, but that it is too challenging to do that). Maybe you could end it with a comment that maybe its not just a PSB selection issue, and that cosmo sims don't capture this quenching process? \sout{Especially if you can connect the PMs that quench}
% \sout{to low gas frac galaxies that were} 
% \\
% \sout{predisposed to quench when PSBs appear to actually} 
% \sout{still have gas e.g. Rowlands et al. 2015; French et al. 2015}}
% \sqcom{I agree with Shoshannah that the conclusions are weak. However, I did not come out with a better interpretation. So, I'll address SB's comment after a co-authors iteration.}. 

To summarise, in this section, we attempt to compare the frequency of quenching in simulated and observed post-mergers. 
All of the three simulations, with the exception of a non statistically robust \eag{} case, produce a much lower excess of rapidly quenching post-mergers compared with observations with a caveat that a robust comparison would need to consider the details of post-starburst galaxy selection versus just rapid quenching.
The incompatibility between observations and simulations may indicate either missing physical ingredients in the simulations, or insufficient resolution to capture the merger-induced effects.
\section{Discussion}
\label{sec:discussion}

% \textcolor{red}{I feel there is room somewhere in the discussion for a small section (two-three paragraphs max) that comments on the agreement and tension between the merger induced AGN activity and quenching (basically there are few highly luminous AGN in post-mergers, in agreement with few rapidly quenching PMs, there are also a small excess of both. BUT the PMs which experience a luminous AGN event are not the ones that quench, instead the quenching appears to be more related to the at coalesence properties of the PMs which is investigated below.). Something short to bring together the two sections above because currently the merger induced AGN activity sections feels a bit unrelated to the rest of the body of work. I can work on this over the weekend, and if I don't have anything substantial by Sunday night we can scrap the idea.}

%%%%%%%%%%%%%%%%%%%%%%%%%%%%%%%%%%%%%%%%%%%%%
%%%%%%%%%%%%%%%%%%%%%%%%%%%%%%%%%%%%%%%%%%%%%
%%%%%%%%%%%%%%%%%%%%%%%%%%%%%%%%%%%%%%%%%%%%%%%%%%

%%%%%%%%%%%%%%%%%%%%%%%%%%%%%%%%%%%%%%%%%%%%%%%%%%%%%%%%%%%%%%%%%%%%%%%%%%%%%%%%%%%%%%%%%%%%%%%%%%%%%%%%%%%%%%%%%%%%%%%%%%%%%%%%%%%%%%%%%%%%%%%%%%%%%%%%%%%%%%%
\subsection{Lack of quenching in simulated post-mergers}
\label{sec:analysis}
%%%%%%%%%%%%%%%%%%%%%%%%%%%%%%%%%%%%%%%%%%%%%%%%%%%%%%%%%%%%%%%%%%%%%%%%%%%%%%%%%%%%%%%%%%%%%%%%%%%%%%%%%%%%%%%%%%%%%%%%%%%%%%%%%%%%%%%%%%%%%%%%%%%%%%%%%%%%%%%
%\begin{figure*}
%\centering
%\setlength\lineskip{-3pt}
%\includegraphics[width=0.335\linewidth]{Plots/EAGLE_%fgas_vs_tpm_wCounts_v000.jpg}
%\hspace{-6pt}
%\includegraphics[width=0.335\linewidth]{Plots/Illust%ris-1_fgas_vs_tpm_wCounts_v000.jpg}
%\hspace{-6pt}
%\includegraphics[width=0.335\linewidth]{Plots/Illust%risTNG100-1_fgas_vs_tpm_wCounts_v000.jpg}
%\\
%\includegraphics[width=0.335\linewidth]{Plots/EAGLE_%fgas_vs_tpm_wQFrac_v000.jpg}
%\hspace{-6pt}
%\includegraphics[width=0.335\linewidth]{Plots/Illust%ris-1_fgas_vs_tpm_wQFrac_v000.jpg}
%\hspace{-6pt}
%\includegraphics[width=0.335\linewidth]{Plots/Illust%risTNG100-1_fgas_vs_tpm_wQFrac_v000.jpg}
%\caption{2D distribution of gas fraction %(f$_\text{gas}$) as a function time after %coalescence (t$_\text{pm}$) of, from top to bottom, %\eag{}, \ill{}, \tng{} simulations. Bins are %colour-coded according to (from left to right) %counts  (i.e., number of galaxies) per bin,  %percentage of quenched post-mergers per bin, and %average black hole mass per bin, respectively. }
%\label{fig:gasFrac}
%\end{figure*}
So far, we have demonstrated that quenching of post-mergers in cosmological simulations occurs rarely and less frequently than in observations. However, we additionally demonstrate that the majority of star forming post-mergers have higher SMBH accretion rates compared with their controls and more frequently host highly accreting SMBHs, which suggests an excess of AGN feedback activity. Here, we investigate pivotal parameters (i.e., M$_\text{BH}$, M$_\text{halo}$, mass ratio $\mu$, and gas fraction f$_\text{gas}$) to get insights into the scarcity of quenching in simulated post-mergers.
Before diving into the analysis, we present a brief overview of the importance of the aforementioned parameters on the interconnection between mergers and quenching. 

Many studies indicate that black hole mass is the best physical proxy for determining the quenching status of massive galaxies, both in simulations \citep[e.g.,][]{Bluck2016, Davies2019, Terrazas2020, Piotrowska2022}, and observations \citep[e.g.,][]{Bluck2016, Terrazas2016, Terrazas2017}. 
In other words, black hole mass directly traces the integrated history of AGN feedback, hence, the cumulative energy released and coupled with the ISM and/or CGM.
Quenched galaxies typically reside in massive halos \citep[e.g.,][]{Davies2019, Cui2021}, harbour overmassive (i.e., massive for their halo mass) black holes and have a low gas fraction, with quenching starting when the total amount of energy from efficient AGN feedback overcomes the gravitational binding energy of the gas \citep[e.g.,][]{Terrazas2020}, and prevents the CGM from
cooling and replenishing the gas reservoir that sustains star formation \citep[e.g.,][]{Davies2020, Zinger2020}.

\citet{Quai2021} empirically found that gas fraction (i.e., the fraction of the total gas over the sum of stellar and gas mass within an aperture of two effective radii) is the best indicator for quenching in IllustrisTNG300-1's galaxies, with star formation halting sharply as soon as gas fraction falls below $0.1$ due to the ejective effect of the TNG kinetic mode feedback.
In this work, we generalise \citet{Quai2021}'s findings to the \eag{}, \ill{} and \tng{} simulations. 
%We confirm that gas fraction is an excellent %parameter for determining the quenching state of %simulated galaxies, though we need to go down to %a
%gas fraction threshold of f$_\text{gas} = 0.05$ %to effectively discriminate
%the quenched from the star-forming population in %\eag{}, \ill{}, and \tng{}.
%Specifically, $97\%$ of galaxies in \ill{} and %$99.7\%$ of galaxies in \tng{} that have %f$_\text{gas} \leq 0.05$ are quenched, whilst %$>99\%$ of galaxies with f$_\text{gas} > 0.05$ %are star-forming.
%In \eag{}, the gas fraction threshold is %slightly less neat than in \tng{}, though still %good at categorizing the star formation state of %the galaxy population, with about $70\%$ of %quenched galaxies having f$_\text{gas} \leq %0.05$, and $95\%$ of star-forming galaxies with %f$_\text{gas} > 0.05$.

It is well established that mass ratio plays an important role in the impact of mergers on star formation \citep[e.g.,][]{Cox2008, Moreno2015, Zheng2020}. About $50\%$ of our post-mergers in the three simulations consist of minor mergers (i.e., $\mu<0.25$, see the bottom-left panel in Figure~\ref{fig:distributions}), hence, it is necessary to tackle whether and how much our results are influenced by mass ratio.
Altogether, the aforementioned parameters allow us to better understand the interplay between mergers, black holes and star formation.

The result of the analysis is shown in Figure~\ref{fig:eag_analysis} for \eag{}, in Figure~\ref{fig:Ill_analysis} for \ill{}, and in Figure~\ref{fig:tng_analysis} for \tng{}.
The figures show the distribution (from top to bottom) of M$_\text{halo}$, gas fraction f$_\text{gas}$, and mass ratio $\mu$ as a function of black hole mass of star-forming (blue points), green valley (green points), and quenching (red points) post-mergers at $500$ Myr after coalescence.
Moreover, the histogram in the top panel of each figure shows the percentage distribution of post-mergers in bins of black hole mass, whilst the three panels on the right show the histograms of the
percentage distribution of post-mergers in bins of (from top to bottom) M$_\text{halo}$, f$_\text{gas}$, and $\mu$.
We discuss the results for each simulation in turn.

%%%%%%%%%%%%%%%%%%%%%%%%%%%%%%%%%%%%%%%%%%%%%%%%%%%%%%%%%%%%%%%%%%
\subsubsection{\eag{}}
\label{sec:eag_lack}
%%%%%%%%%%%%%%%%%%%%%%%%%%%%%%%%%%%%%%%%%%%%%%%%%%%%%%%%%%%%%%%%%%
\begin{figure}
\centering
\includegraphics[width=\linewidth]{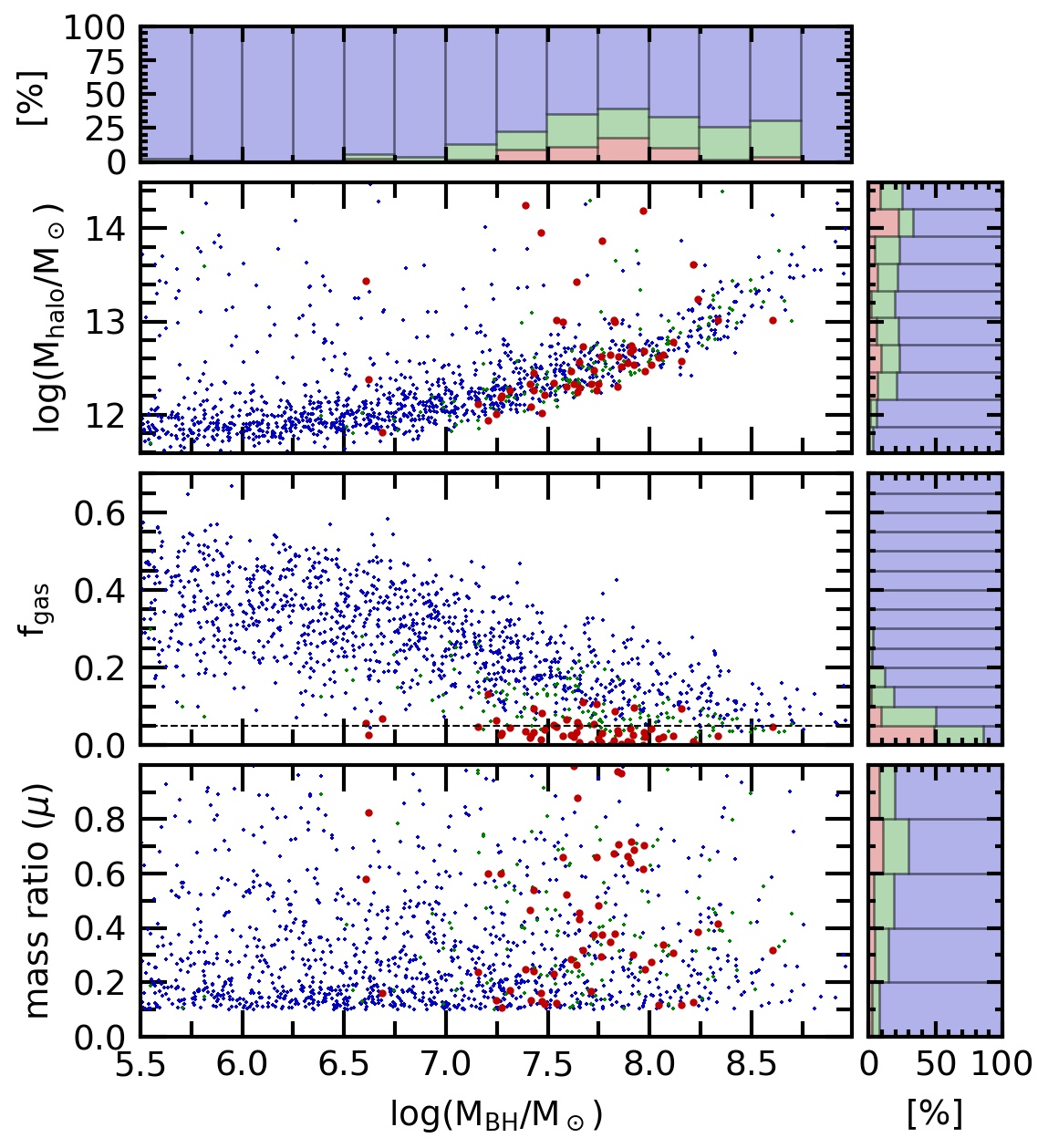}\\
\caption{The three main panels show, from top to bottom, the halo mass (log(M$_\text{halo}/$M$_\odot$)), the gas fraction (f$_\text{gas}$), and the mass ratio ($\mu$) as a function of the black hole mass (log(M$_\text{BH}/$M$_\odot$)) for \eag{} post-mergers at $500$ Myr after the merger.  
Star-forming post-mergers are represented in blue, green valley post-mergers are represented in green and quenched post-mergers in red. 
The histogram above the top panel shows the percentage distribution of star-forming, green valley and quenched post-mergers in bins of black hole mass. The histograms at the right side of the main panels represent the percentage distributions of post-mergers in bins of, from top to bottom, halo mass, gas fraction, and mass ratio. 
The figure shows that the vast majority of \eag{} post-mergers do not quench, mostly because they have small black holes, but also because some of them reside in massive halos in which the AGN feedback is not able to interrupt the cooling of dense gas in the centre of their circumgalactic medium \protect\citep[read Appendix A in][]{Davies2019}.
}
\label{fig:eag_analysis}
\end{figure}
We discuss the results starting from the analysis of \eag{} in Figure~\ref{fig:eag_analysis}. 
The histogram in the top panel demonstrates that quenching in \eag{} post-mergers is rare in any black hole mass regime. 
Almost all quenched post-mergers have black hole masses between $10^{7.1}$ and $10^{8.25}$ M$_\odot$, although, this mass regime is dominated by star-forming post-mergers, representing $60-80$ percent of the population.
At lower black hole mass, we find a long tail of $740$ (i.e., the $55\%$ of \eag{} post-mergers) galaxies, out of which only $3$ and $18$ are quenched and green valley post-mergers, respectively.
However, at black hole masses larger than $10^{8.25}$ M$_\odot$,
where we would expect the quenched population to prevail, we find that $73\%$ of post-mergers are star-forming, and only $2$ out of $94$ are quenched.

To better understand why \eag{} post-mergers do not quench promptly, we investigate the M$_\text{halo}$ - M$_\text{BH}$ and f$_\text{gas}$ - M$_\text{BH}$ relationships (top and central distributions, respectively, in Figure~\ref{fig:eag_analysis}).
All post-mergers with less massive black holes reside in small halos and have gas fractions large enough to keep sustaining new episodes of star formation.
In this black hole mass regime, it is clear that AGN feedback is not effective\footnote{The modified Bondi-Hoyle accretion implemented in \eag{} is proportional to M$_\text{BH}$ \citep[][]{RosasGuevara2015, RosasGuevara2016}.} in removing gas from the ISM, and the low virial temperature and high CGM mass fraction (and hence characteristic density of the CGM) of the halos grants short cooling time, thus providing fuel for star formation.
It is less straightforward to interpret the dearth of quenching in post-mergers with black holes more massive than $10^{8.25}$ M$_\odot$.
The extra complication arises because \eag{}'s AGN feedback is not efficient enough at ejecting baryons and preventing cooling in group and cluster haloes. 
This is the reason why the \eag{} model produces groups and clusters that are too bright in X-ray \citep[see Appendix A in ][]{Davies2019, Schaye2015}.
The CGM mass fraction is high in these haloes, hence, their inner regions can become quite dense, giving a short cooling time \citep[see][]{Davies2019, Oppenheimer2020}.
It is then very difficult for the AGN to remove gas from the CGM, which can therefore replenish the galaxy supply for new star formation.

We can observe the effect of change in AGN ``effectiveness'' by looking at the f$_\text{gas}$-M$_\text{BH}$ distribution, and at the histogram of the percentage distribution of post-mergers in bins of f$_\text{gas}$ in the central panels of Figure~\ref{fig:eag_analysis}.
The panels show a trend with f$_\text{gas}$ reducing with increasing M$_\text{BH}$ (and increasing M$_\text{halo}$).
However, the vast majority of post-mergers maintain f$_\text{gas} > 0.05$, thus being able to form stars and stay in the star-forming main sequence.
The few post-mergers that quench within $500$ Myr after coalescence have gas fraction around or below $0.05$.

In the bottom panel of Figure~\ref{fig:eag_analysis}, we investigate how the merger mass ratio influences the quenched fraction in \eag{} post-mergers. 
We find that major mergers show a slightly larger chance to be quenched, but even at mass ratios $\mu>0.7$, only a small fraction of $7.3\%$ post-mergers are quenched, to which can be added $15.3\%$ of post-mergers in the green valley.

The weak dependence of quenched fraction on merger mass ratio is in contrast with some other results in the literature.
For instance, \citet{Davies2022} found that increasing mass ratio in a \eag{} Milky Way-like merger galaxy causes more violent and preferentially head-on collisions, maximising the probability of quenching star formation.
If we limit the analysis to Milky Way-like post-mergers with $12.5 \leq$ log(M$_\text{halo}/$M$_\odot) \leq 12.7$, we find that major mergers ($\mu >0.5$) have around $20\%$ higher chance to quench (or to reside in the green valley) shortly after coalescence than in minor mergers.  Although, we also find that as much as $63\%$ of galaxies in this regime (i.e., major mergers in Milky Way-like galaxies) keep to stay star-forming.

To summarise, in \eag{} we find that as halo mass increases, the chances of a galaxy to be quenched by AGN feedback increases up to M$_\text{BH} \sim 10^{8.2}$ M$_\odot$, and then decreases as the AGN stop being effective in massive halos.
Moreover, we do not find a significant impact of the merger mass ratio on the quenched fraction of \eag{} post-merger galaxies. 
Ultimately, we find that the gas fraction is the key ingredient to clearly disentangle star-forming and quenched post-mergers.

%%%%%%%%
%%%%%%%%%%%%%%%%%%%%%%%%%%%%%%%%%%%%%%%%%%%%%%%%%%%%%%%%%%%%%%%%%%
\subsubsection{\ill{}}
\label{sec:ill_lack}
%%%%%%%%%%%%%%%%%%%%%%%%%%%%%%%%%%%%%%%%%%%%%%%%%%%%%%%%%%%%%%%%%%
\begin{figure}
\centering
\includegraphics[width=\linewidth]{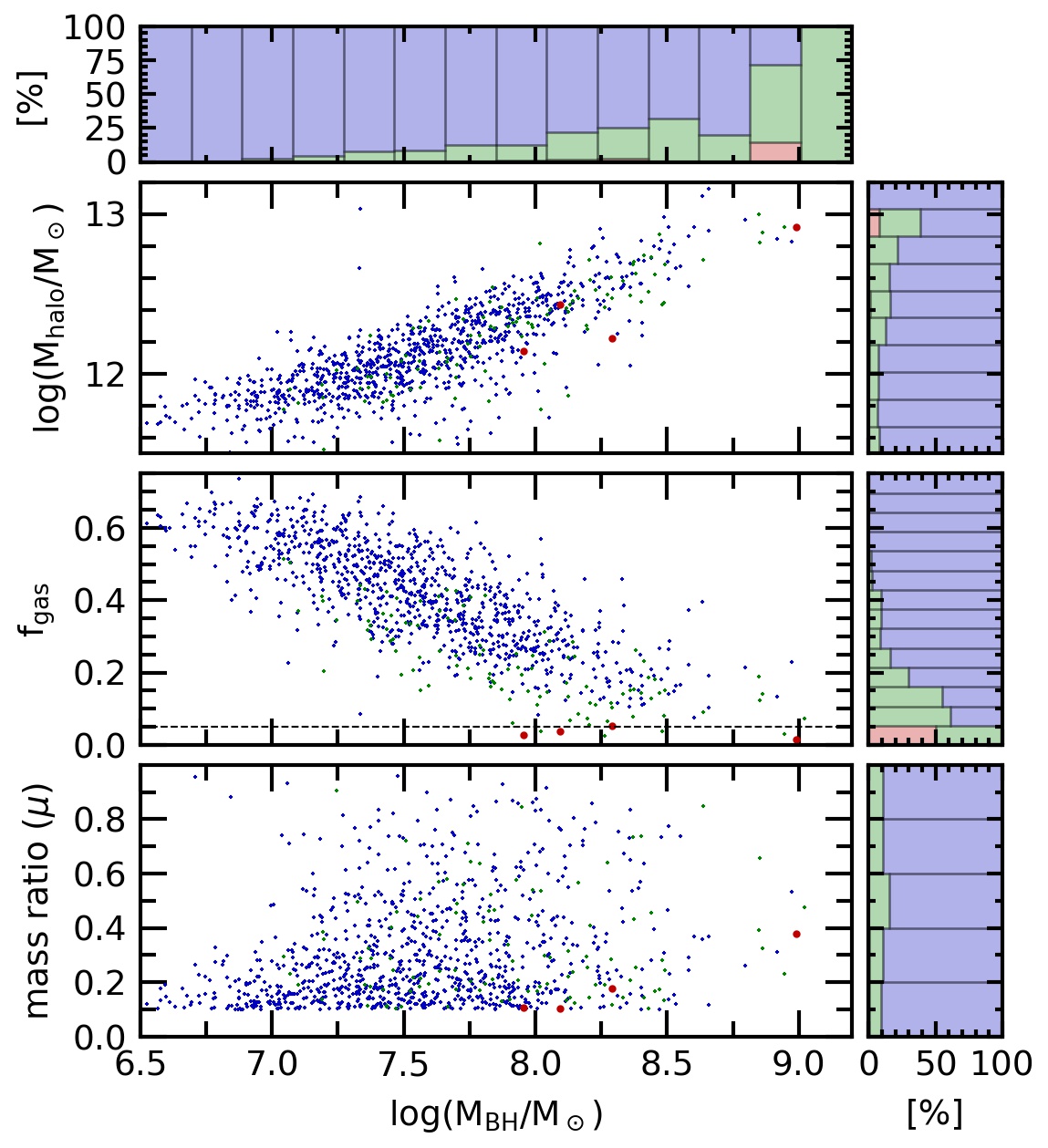}\\
\caption{Same as in Figure~\protect\ref{fig:eag_analysis}, but for \ill{}. Note that the range in black hole masses differs from that in Figure~\protect\ref{fig:eag_analysis}, due to the difference in black hole mass distribution between \eag{} and \ill{} post-mergers (see Figure~\protect\ref{fig:distributions}). 
The figure shows that only four \ill{} post-mergers 
quench star formation within $500$ Myr after coalescence. The fraction of green valley post-mergers increases with increasing halo and black hole mass. However, quenching happens only in the few galaxies showing a gas fraction as low as $0.05$ or less. The mass ratio does not affect the quenching fraction. Indeed, the fraction of green valley galaxies does not increase with the increasing mass ratio.}
\label{fig:Ill_analysis}
\end{figure}

%\textcolor{red}{Overall comment %that maybe this section should be %subsectioned by simulation, since %there is a lot of text and many %ideas to break down its easy to %get lost.}

Figure~\ref{fig:Ill_analysis} shows the results regarding \ill{}. 
The figure demonstrates that only four \ill{} post-mergers quench within $500$ Myr after the merger. Despite the fact that the gas fraction decreases steadily with increasing black hole mass (and increasing halo mass), the star formation is affected (there is an increasing number of green valley galaxies), though not completely halted.
The implementation of the \ill{} AGN feedback is the reason for such complete lack of quenching in post-mergers. Indeed, \ill{} post-mergers show a strong black hole accretion enhancement (see Section~\ref{sec:accretion}), thus activating the high accretion mode of the AGN feedback, which deposits thermal energy with a low coupling efficiency \citep[e.g.,][]{Sparre2015, Donnari2019, Bluck2016, Terrazas2017}, leaving the surrounding ISM almost unaffected.

%%%%%%%%%%%%%%%%%%%%%%%%%%%%%%%%%%%%%%%%%%%%%%%%%%%%%%%%%%%%%%%%%%
\subsubsection{\tng{}}
\label{sec:tng_lack}
\begin{figure}
\centering
\includegraphics[width=\linewidth]{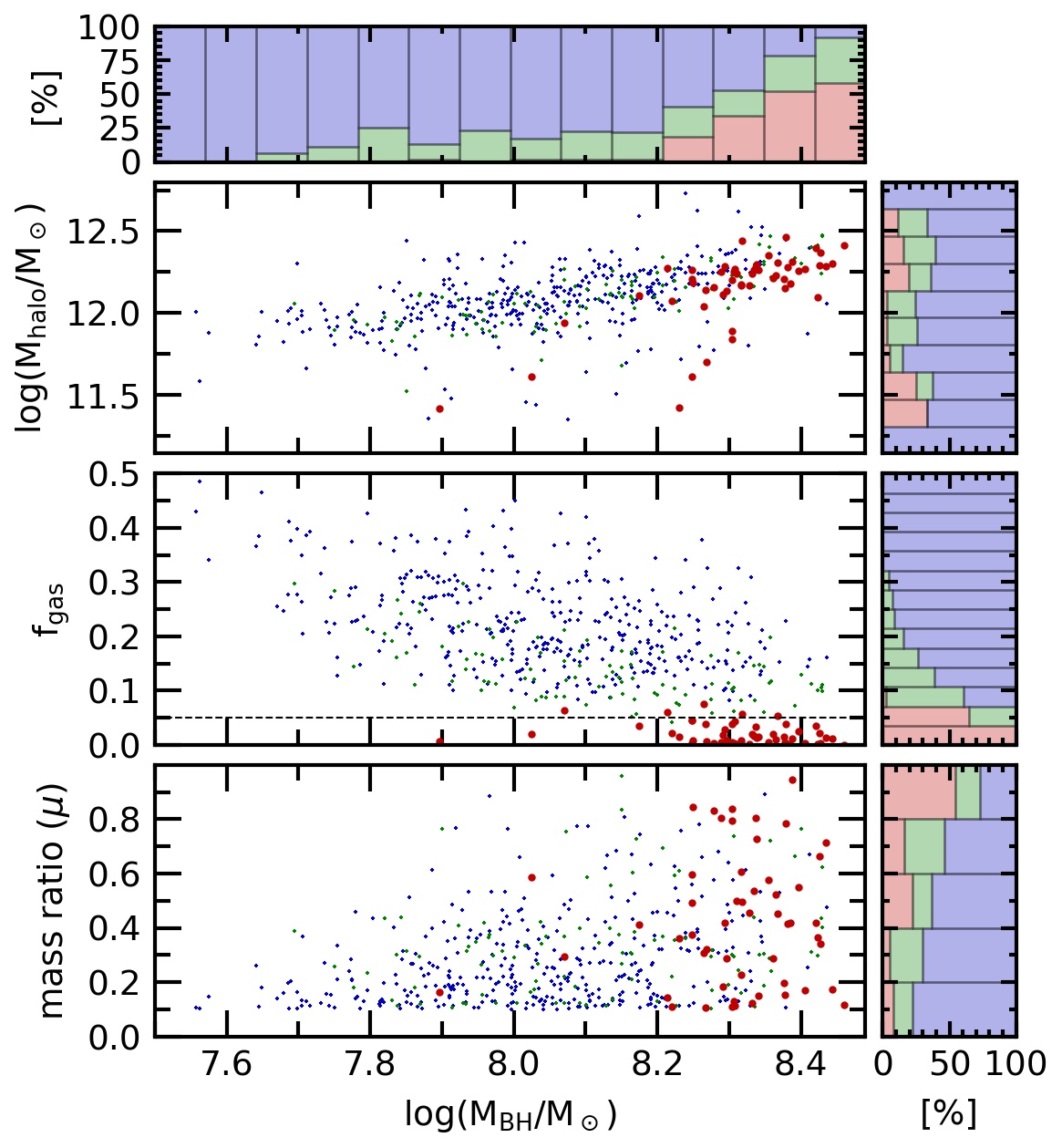}\\
\caption{Same as in Figure~\protect\ref{fig:eag_analysis}, but for \tng{}. Note that the range in black hole masses differs from that in Figure~\protect\ref{fig:eag_analysis}, due to the difference in black hole mass distribution between \eag{} and \tng{} post-mergers (see Figure~\protect\ref{fig:distributions}).
The figure shows that the importance of quenching in \tng{} post-mergers starts arising at black hole masses above $10^{8.2}$ M$_\odot$, the black hole mass threshold above which
more than $90\%$ of the IllustrisTNG galaxies are quenched \protect\citep[e.g.][]{Terrazas2020, Zinger2020}. However, quenching occurs only in those post-mergers with massive black holes that have also a very low gas fraction, f$_\text{gas}\leq0.05$ \protect\citep[see also][]{Quai2021}. The figure shows also that major mergers have a larger chance to quench star formation.}
\label{fig:tng_analysis}
\end{figure}

%%%%%%%%%%%%%%%%%%%%%%%%%%%%%%%%%%%%%%%%%%%%%%%%%%%%%%%%%%%%%%%%%%
We focus now on the analysis of the lack of quenching in \tng{}'s post-mergers. 
The results are shown in Figure~\ref{fig:tng_analysis}, with an identical layout as that in Figure~\ref{fig:eag_analysis} for \eag{}.
Studies of quenching in IllustrisTNG \citep{Weinberger2017, Nelson2018, Terrazas2020} demonstrated that only the kinetic mode of AGN feedback is capable of causing star formation suppression.
\citet{Terrazas2020} showed that the ejective feedback in IllustrisTNG becomes effective at quenching star formation once the cumulative kinetic energy overcomes the total gravitational binding energy of the gas in a galaxy. 
They also demonstrated that the kinetic feedback process dominates in galaxies whose M$_\text{BH}$ exceeds $10^{8.2}$ M$_\odot$, that is the black hole mass threshold above which more than $90\%$ of the IllustrisTNG galaxies are quenched \citep[see also ][]{Zinger2020}. 
Such a M$_\text{BH}$ threshold for quiescence arises from the IllustrisTNG model parameters chosen to reproduce observational properties of the galaxy population at present \citep{Pillepich2018a}. 
Our results concerning the quenching deficiency in the \tng{} post-merger population reflect the general quenching process of galaxies in the simulation.
In detail, the histogram in the top panel of Figure~\ref{fig:tng_analysis} shows that almost none of the post-mergers with M$_\text{BH} < 10^{8.2}$ M$_\odot$ are quenching, whilst the percentage of quenched post-mergers increases steadily at larger black hole mass, passing from $18\%$ at M$_\text{BH} \sim 10^{8.3}$ M$_\odot$, to $58\%$ at M$_\text{BH} \sim 10^{8.4}$ M$_\odot$, to which we can add, respectively, $22\%$ and $33\%$ of post-mergers in the green valley.
All TNG quenched post-mergers have gas fraction f$_\text{gas} \leq 0.05$ (see the central panels in Figure~\ref{fig:tng_analysis} ), confirming that these
galaxies ran out of fuel for star formation. 
Differently from \eag{} though, we do not find trends with M$_\text{halo}$ (see the top panels in Figure~\ref{fig:tng_analysis}).
However, our \tng{} post-merger population has a smaller mass dynamical range, with most of the massive galaxies residing in
halos of around $10^{12.5}$ M$_\odot$, hence, we cannot probe the quenching efficiency of IllustrisTNG AGN feedback in post-mergers located in groups and clusters, as we can do for the \eag{} sample.

In the bottom panels of Figure~\ref{fig:tng_analysis}, we investigate whether merger mass ratio influences the quenched fraction in \tng{} post-mergers.
The histogram in the bottom-right corner shows that the percentage of quenched post-mergers strongly increases with increasing mass ratio, with a peak of $55\%$ quenched post-mergers in the bin of mass ratio above $0.8$. 
However, the distribution of mass ratio as a function of black hole mass (bottom-left panel of Figure~\ref{fig:tng_analysis}) shows a trend
where higher mass ratio mergers have larger black hole mass, thus with a higher AGN feedback efficiency.
To test whether the increment of the quenched fraction with the mass ratio is attributable to a larger black hole population, we calculate the fraction of quenched galaxies in two bins of mass ratio (i.e., $\mu \leq 0.5$, and $\mu> 0.5$) for a sub-sample of \tng{} post-mergers that have black holes more massive than $10^{8.2}$ M$_\odot$.
In this sub-sample, we find $28.4\%$ and $41.9\%$ of quenched
post-mergers in the minor and major merger bins, respectively, thus suggesting that \tng{} major mergers increase the chance of quenching shortly after coalescence.
However, we find that as much as $39.5\%$ of major mergers that exhibit a massive black hole continue to reside in the star-forming main sequence. Such lack of quenching in major mergers could be due to other parameters governing the evolution in post-merger phases. 
For instance, \citet{Zeng2021} showed that progenitor orbits have an important effect on the remnant of \tng{} post mergers, with prograde orbits resulting in post-mergers with a high chance to preserve the original disk and a star-forming configuration even in major merger events. 

To summarise, we find that quenching in a \tng{} post-merger occurs promptly after coalescence only if its black hole mass is larger than $\sim10^{8.2}$ M$_\odot$ \citep[i.e., the mass above which the \tng{}'s AGN feedback switches to the effective kinetic mode, ][]{Weinberger2018, Terrazas2020}, and at the same time, its gas fraction is below $0.05$. 
We also find that the \tng{} quenching fraction increases with increasing mass ratio, even though around $40$ percent of major mergers harbouring massive black holes continue to be star-forming long after coalescence. 

%%%%%%%%%

%%%%%%%%%%%%%%%%%%%%%%%%%%%%%%%%%%%%%%%%%%%%%%%%%%%%%%%%%%%%%%%%%%%
%%%%%%%%%%%%%%%%%%%%%%%%%%%%%%%%%%%%%%%%%%%%%%%%%%%%%%%%%%%%%%%%%%%%%%%%%%%%%%%%%%%%%%%%%%%%%%%%%%%%%%%%%%%%%%%%%%%%%%%%%%%%%%%%%%%%%%%%%%%%%%%%%%%%%%%%%%%%%%%
\subsection{Gas loss}
\label{sec:gas_analysis}
%%%%%%%%%%%%%%%%%%%%%%%%%%%%%%%%%%%%%%%%%%%%%%%%%%%%%%%%%%%%%%%%%%%%%%%%%%%%%%%%%%%%%%%%%%%%%%%%%%%%%%%%%%%%%%%%%%%%%%%%%%%%%%%%%%%%%%%%%%%%%%%%%%%%%%%%%%%%%%%
\begin{figure}
\centering
\includegraphics[width=\linewidth]{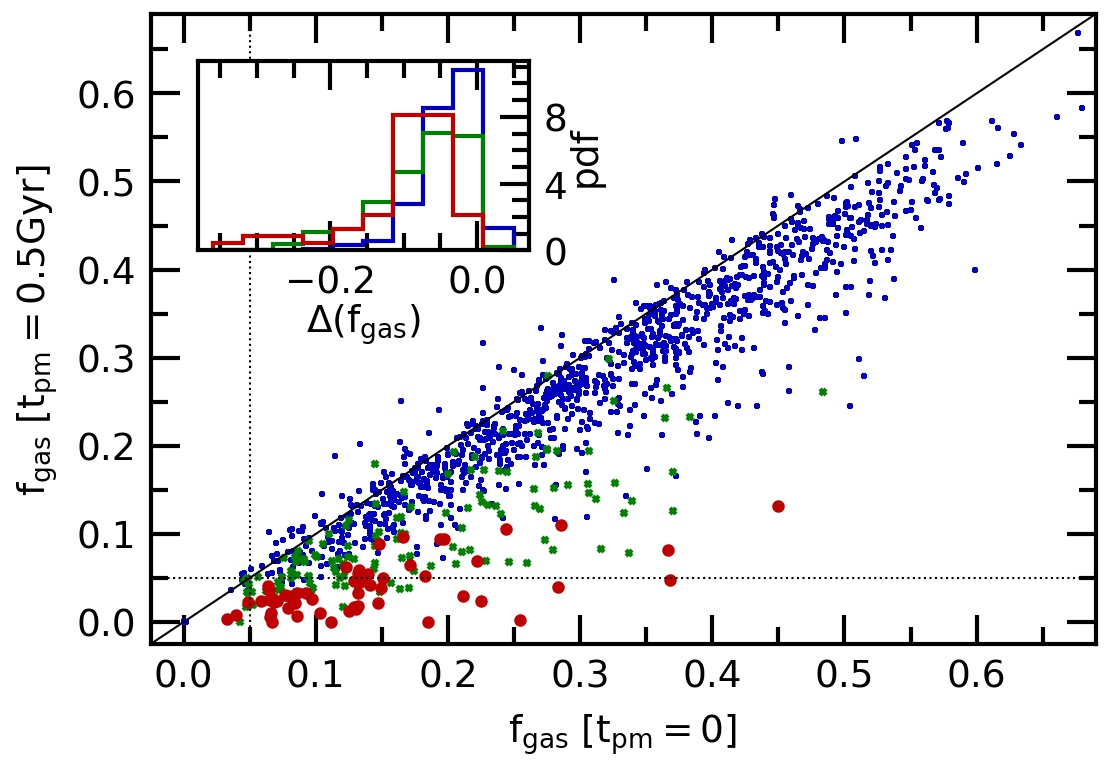}\\
\includegraphics[width=\linewidth]{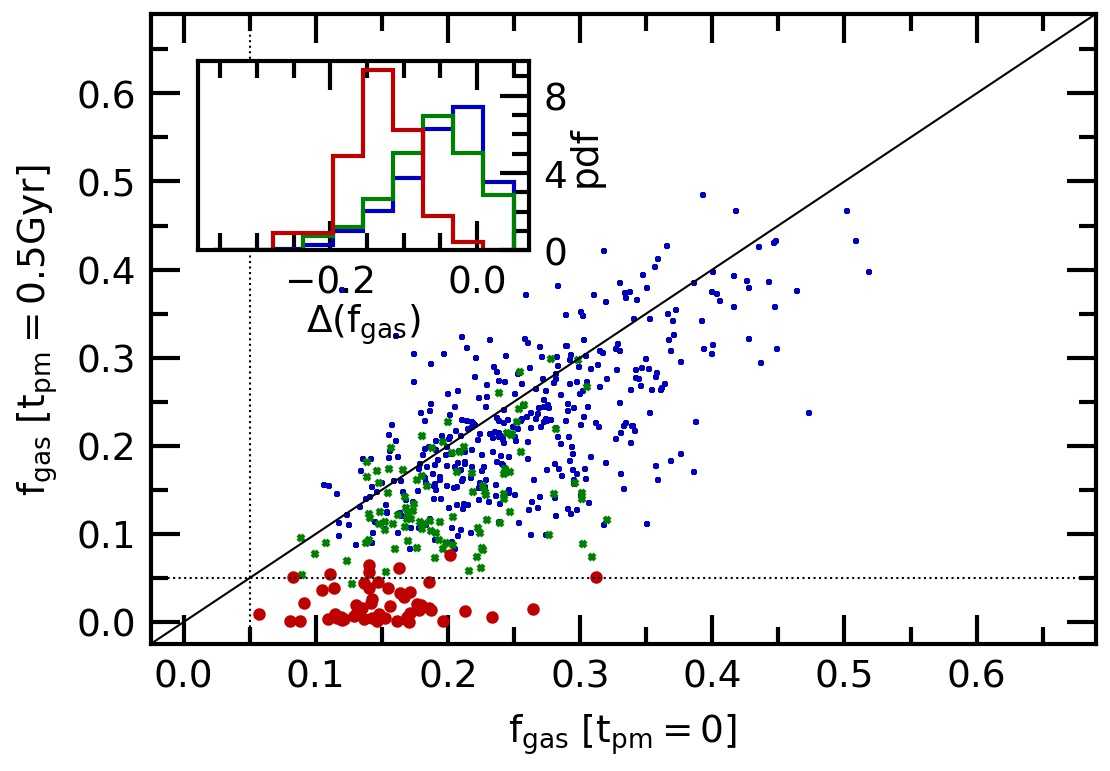}\\
\caption{Distribution of gas fraction at $500$ Myr after the merger versus the gas fraction at coalescence. 
The top panel shows the result for the \eag{} simulation, whilst the bottom panel is for the \tng{} simulation.
Star-forming post-mergers are represented in blue, green valley post-mergers in green, whilst quenched post-mergers in red.
The insets in both panels represent the distribution of $\Delta \text{f}_\text{gas}$ (i.e., f$_\text{gas}$ [t$_\text{pm} = 0.5$ Gyr] $-$ f$_\text{gas}$ [t$_\text{pm} = 0$]).
The panels show that two things are necessary to quench post-merger galaxies in both simulations : (1) a predisposition towards quenching (i.e., most of the quenched post-mergers have small gas fraction to start with), and (2) experience a large amount of gas loss shortly after the merger.
}
\label{fig:gas_loss}
\end{figure}
In the previous sections, we showed that most simulated post-mergers keep forming stars. 
However, even if quenching is less frequent than in observations, we find that post-mergers quench at circa twice and at around ten times the rate of non-interacting control galaxies, in \tng{} and  \eag{}, respectively (see Figure~\ref{fig:excess}). 

In the previous section the role of effective AGN feedback in quenched \eag{} and \tng{} post-mergers was investigated, a process which is in common with non-interacting galaxies. 
However, since we matched the two populations (and kept the match over the entire simulation life) in black hole mass (see Section~\ref{sec:control_sample}), the quenching excess cannot be completely attributable to a mere difference in the black hole population, hence, it could be related to the merger process itself. 
In this section, we aim to tackle the origin of the extra quenching in \eag{} and \tng{} post-mergers. 
Note that we do not further analyse \ill{}'s post-mergers because in Section~\ref{sec:fractions} we demonstrated that quenched post-mergers in \ill{} are rarer than what expected from control non-interacting galaxies.

In Figures~\ref{fig:eag_analysis}, \ref{fig:Ill_analysis}, \ref{fig:tng_analysis}, it emerged that gas fraction is an excellent discriminator for the star-formation status of a population. 
Particularly, we found that a low level of gas fraction (i.e., f$_\text{gas} \leq 0.05$) strongly correlates with quenching. 
To shed light on the surplus of quenching in post-mergers compared to their direct control galaxies, it is, therefore, worth investigating how gas fraction evolves from coalescence in the post-merger population.
We present the results in Figure~\ref{fig:gas_loss}. 
The two panels show the gas fraction at $500$ Myr after the merger versus the gas fraction at coalescence for (from top to bottom) \eag{} and \tng{}.
Star-forming post-mergers are represented in blue, green valley post-mergers in green and quenched post-mergers in red.
The inset in each panel shows the distributions of gas loss, $\Delta \text{f}_\text{gas}$ (i.e., f$_\text{gas}$ [t$_\text{pm} = 0.5$ Gyr] $-$ f$_\text{gas}$ [t$_\text{pm} = 0$]), of the three star formation categories (i.e., star-forming, green valley, and quenched).
We find two key results from the analysis of Figure~\ref{fig:gas_loss}. 
First of all, the plot shows that $82.5\%$ and $91\%$ (in \eag{} and \tng{}, respectively) of quenched post-mergers had low gas fractions at coalescence\footnote{We stress that at coalescence all the galaxies in our samples have to be star-forming by construction, see Section~\ref{sec:data}.}, with values f$_\text{gas}\leq0.2$. 
Such low gas fraction values at the time of the merger suggest that there is a pre-disposition towards quenching when the gas fraction is low to start with \citep[see also][]{Quai2021}.
What is also interesting is that the distributions of $\Delta \text{f}_\text{gas}$ shown in the insets of Figure~\ref{fig:gas_loss} demonstrate that quenched post-mergers in both simulations are also experiencing ``preferentially'' larger gas loss compared to those that stay star-forming and in the green valley. 
In numbers, we find mean values of $\Delta \text{f}_\text{gas}=-0.1$ and $=-0.13$ for \eag{} and \tng{} quenched post-mergers, respectively. Conversely, we find mean values of $\Delta \text{f}_\text{gas}=-0.03$ for both \eag{} and \tng{} star-forming, and $\Delta \text{f}_\text{gas}=-0.07$ and $=-0.06$ for \eag{} and \tng{} green valley post-mergers.
Therefore, on the one hand, there is a ``pre-disposition'' to quench. However, on the other hand, the quenched post-mergers tend to lose relatively more gas shortly after coalescence. 

\begin{figure}
\centering
\includegraphics[width=\linewidth]{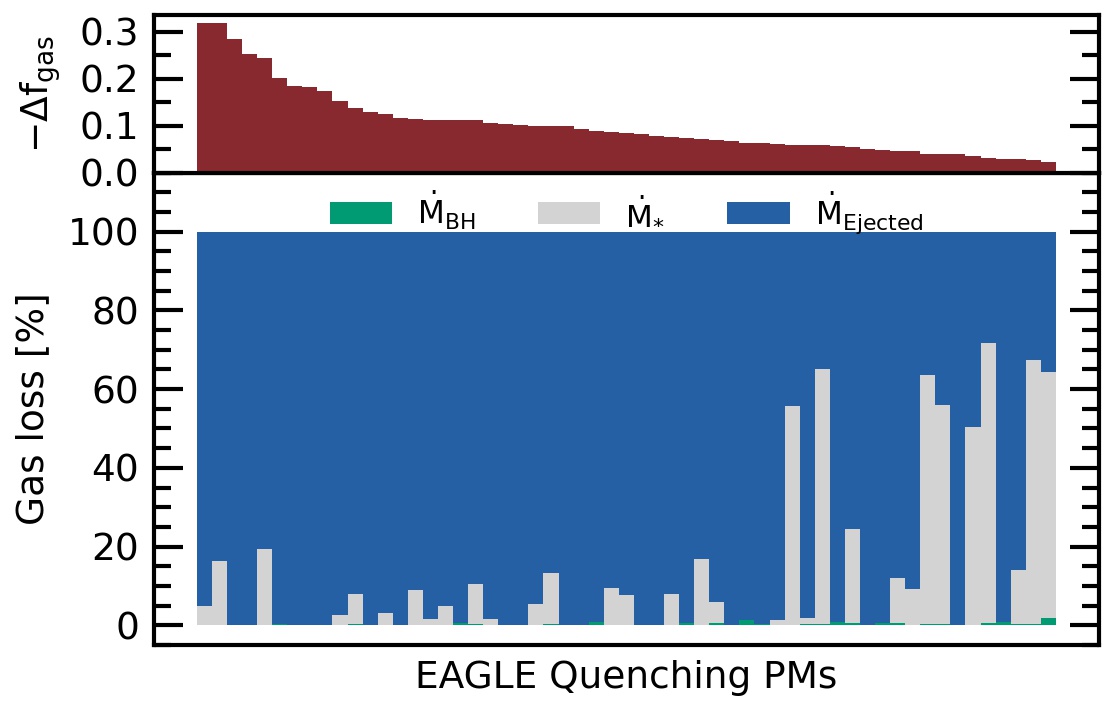}
\includegraphics[width=\linewidth]{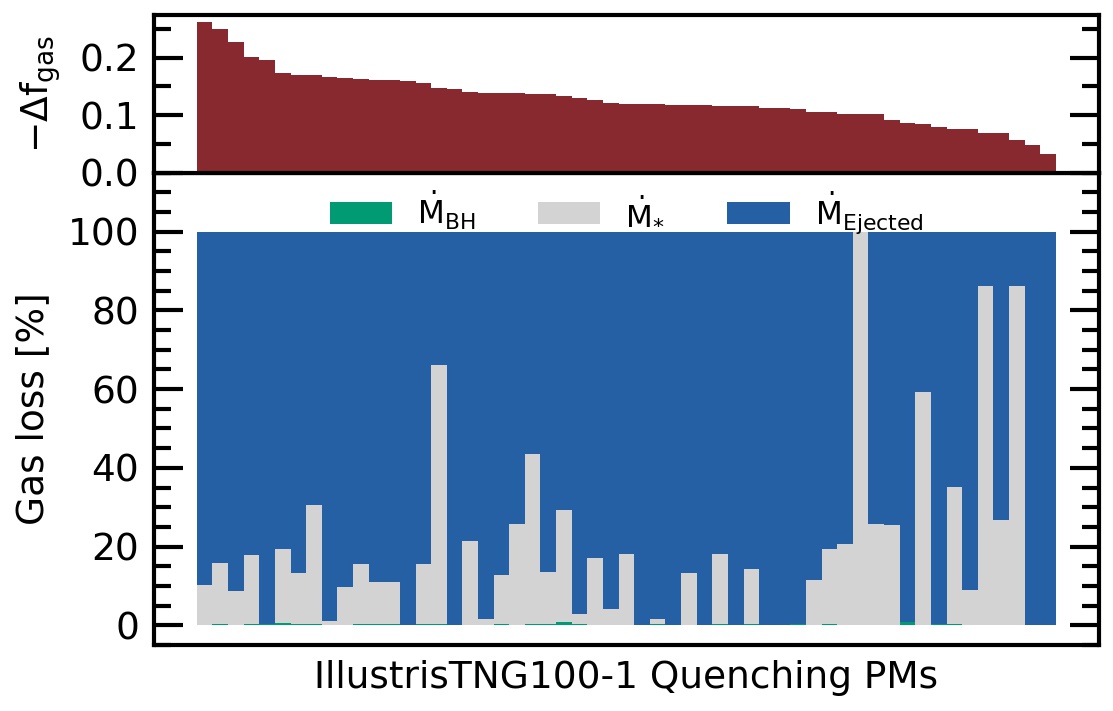}\\
\caption{In the top panel, each vertical line represents a single \eag{} quenched post-merger, whilst in the bottom panel, each vertical line represents a single \tng{} quenched post-merger.
Galaxies are sorted from right to left from lowest to highest $-\Delta \text{f}_\text{gas}$ values (dark red distribution in the top of both panels), which means according to the increasing amount of gas missing. 
Then, for each galaxy we split the percentage of gas loss into three groups: (1) gas that has been accreted onto the central black hole (green bar), (2) gas that has been converted into new stars (grey bar), and (3) gas that has been pushed out of the galaxy (blue bar).}
\label{fig:gas_loss_hist}
\end{figure}
We investigate further the latter hypothesis in Figure~\ref{fig:gas_loss_hist}, as well as attempting to quantify the mechanism responsible for the gas depletion that leads to quenching.
Each vertical bar represents a single post-merger that quenches within $500$ Myr after the merger. In the top panel we focus on \eag{}, while quenched post-mergers in \tng{} are presented in the bottom panel.
Galaxies are sorted from right to left from lowest to highest $-\Delta \text{f}_\text{gas}$ values (dark red histogram-like distribution), or in other words, according to increasing amount of gas missing with respect to coalescence. 
For each galaxy, we then show the percentage of such gas that has been accreted onto the central black hole (green bar; almost invisible for most galaxies), that has been converted into new stars (grey bar), and the percentage of gas that has been ejected outside the galaxy (blue bar).  
Figure~\ref{fig:gas_loss_hist} shows that in both simulations there is a large predominance of post-mergers which have $>80-100\%$ of their missing gas completely ejected out of the galaxy, especially at higher $-\Delta \text{f}_\text{gas}$ values.

%(1) all \eag{} and $80\%$ of \tng{} post-mergers %that have consumed more than their $50\%$ of %initial gas to form new stars have %preferentially smaller $-\Delta \text{f}_\text{gas}$ %values, that means low gas fraction to start %with, or what we have previously called %``pre-disposition'' towards quenching; and (2) %with increasing 
In summary, in this section we uncovered the two most common scenarios through which post-mergers quench following a merger event:  galaxies which are predisposed to quench due to low gas fractions and those that quench after a period of significant gas evacuation.

%\sqcom{Since we matched post-mergers and controls in black hole mass,  the excess of quenching compared to the non-interacting population can only be ascribed to some dynamical effect of the merger itself. However, the physical mechanism that contributes to rapidly shutting down star formation in post-coalescence galaxies is not further addressed by our analysis.} ###same comment as above with the BH mass matching. I'm not sure I agree, yes the integrated feedback history is the same but I don't think recent rapid growth vs. early growth with a virtually non-accreting BH today - resulting in the same BH mass - rules out AGN feedback variation as a quenching mechanism###

%\begin{figure}
%\centering
%\includegraphics[width=\linewidth]{Plots/EAGLE_HistGasLoss_WMassRatio_v400.jpg}
%%\includegraphics[width=\linewidth]{Plots/Illustris-1_HistGasLoss_WMassRatio_v400.jpg}
%\includegraphics[width=\linewidth]{Plots/IllustrisTNG100-1_HistGasLoss_WMassRatio_v400.jpg}\\
%\caption{}
%\label{fig:gas_loss}
%\end{figure}

%%%%%%%%%%%%%%%%%%%%%%%%%%%%%%%%%%%%%%%%%%%%%%%%%%%%%%%%%%%%%%%
%%%%%%%%%%%%%%%%%%%%%%%%%%%%%%%%%%%%%%%%%%%%%%%%%%%%%%%%%%%%%%%
\subsection{Rapid versus long-term quenching}
\label{sec:rapid_vs_long}
%%%%%%%%%%%%%%%%%%%%%%%%%%%%%%%%%%%%%%%%%%%%%%%%%%%%%%%%%%%%%%%
%%%%%%%%%%%%%%%%%%%%%%%%%%%%%%%%%%%%%%%%%%%%%%%%%%%%%%%%%%%%%%%
In this work we probed a merger-driven quenching scenario within state-of-the-art hydrodynamical cosmological simulations that encompass complete and robust samples of mergers covering a wide range of halo mass, mass ratio, orbital parameters, physical properties and environment. 
We found that a prompt quenching following a merger event is much less frequent than that expected from observational results and previous generations of simulations \citep[see also][for similar results]{Weinberger2018, RodriguezMontero2019, Pathak2021}.
We limited our analysis to the early stages of the post-merger evolution (i.e., probing the causal connection between mergers and rapid quenching) for two reasons: (1) the quenched fraction in post-mergers tends to gradually level off and finally even out with the quenched fraction of the control population (i.e., post-merger evolution becomes statistically indistinguishable from the secular evolution of non-interacting galaxies, see Figures~\ref{fig:fractions} and \ref{fig:excess}); and (2) observations of quenched post-mergers are confined within a few hundred Myr after the coalescence phase.  
However, we cannot exclude the possibility of a long-term quenching that might be causally connected to galaxy-galaxy interactions. 
For instance, \citet{Davies2022} showed that in a series of genetically modified mergers of Milky Way-like galaxies from the \eag{} simulation, the intense AGN feedback triggered by major mergers can modify the evolution of the galaxy-CGM ecosystem. In such controlled post-mergers, the AGN feedback can eject enough gas to promptly push the galaxy out of the main sequence but not to fully quench its star formation. 
Quenching arises typically only around $3$ Gyr after the merger, due to the prolonged CGM cooling time that prevents further inflow of new gas. 
Testing longer-term quenching observationally is much more of a challenge as it requires being able to time the quenching episode as well as timing merger features (to prove a causal connection between them).  Whilst star formation histories might be able to give fairly reliable quenching times in the last few Gyrs, timing mergers is harder, although some machine learning applications are now attempting to do this \citep[e.g.,][Bottrell et al., in preparation]{Bottrell2022}.

\section{Summary and Conclusions}
\label{sec:conclusions}
We have used three different cosmological simulations to analyse a complete sample of post-merger galaxies in a fully cosmological environment spanning a diverse range of merger histories and physical properties.
Each simulation implements accretion onto the central supermassive black hole and AGN feedback models that are distinct (see, for instance, Section~\ref{sec:data}).
%All the simulations are calibrated to %fit fundamental global scale relations %in the local Universe, via a tuning of %specific parameters. 
%For example, \eag{} has been calibrated %to reproduce the mass function and the %M$_\text{BH}$-M$_\text{bulge}$ relation %(assuming M$_\text{bulge}=$M$_\ast$) at %present time, as well as reasonable %galaxy sizes \citep{Crain2015, %Schaye2015}, whereas IllustrisTNG was %calibrated based on the cosmic star %formation rate density, the galaxy mass %function, the stellar-to-halo ratios, %and the M$_\text{BH}$-M$_\ast$ relation %\citep{Pillepich2018a}. 
%Calibrated models allow a better %agreement with the global galactic %population, however, there could persist %some residual non negligible differences %in the impact of different sub-grid %models on galaxy evolution that often %lead to unalike predictions of physical %processes \citep[e.g.,][]{Somerville2015%, Schaye2015, Weinberger2017}.

Although galaxy-galaxy mergers are rare events in the history of a galaxy, there is a general consensus  that mergers have a tremendous impact on the evolution of the merger remnants, with repercussions on shaping the global properties
of galaxies (e.g., dark and stellar mass increment, metallicity evolution, mass function evolution, etc.). 
Such relative importance of mergers in the bigger picture of galaxy evolution should naturally arise as predictions of the models implemented in simulations.
In this paper, we quantify the effect of galaxy-galaxy mergers on the black hole accretion, AGN feedback and rapid star formation quenching in three state-of-the-art cosmological simulations: \eag{}, \ill{}, and \tng{}. 
Moreover, for the first time we can compare the incidence of quenching in simulations with that of a large and pure sample of observed post-merger galaxies analysed in \citet{Ellison2022}.
Our findings can be summarised as follows:

\begin{itemize}
    \item The majority of star forming post-mergers exhibit higher SMBH accretion rates when compared with matched controls. On average, the post-merger population exhibits SMBH accretion rates 2.5 (\eag{}), 4.5 (\ill{}), and 2.1 (\tng{}) times higher than matched controls (see Section~\ref{sec:accretion}, and Figures~\ref{fig:SMBHARenhancements} and~\ref{fig:fracAGNEvent}. %Significantly, not all post-mergers have enhanced SMBH accretion rates, with 20-35\% exhibiting accretion rates below matched controls.
    
    \medskip
    
    \item Highly accreting SMBHs occur more frequently in the post-merger sample compared to the control sample, where post-mergers host the highest SMBH accretion rates up to 4 times more often (see Figure~\ref{fig:fracAGNExcess}).
    
    \medskip
    
    \item Rapid quenching is rare in the post-merger population in all three cosmological simulations. Only 4.5\% (\eag{}), 0.4\% (\ill{}), and 10\% (\tng{}) of post-mergers quench within 500 Myrs of coalescence (see Section~\ref{sec:fractions} and Figure~\ref{fig:fractions}). Despite the rarity of quenched post-mergers, quenching occurs more frequently than in the control population in the \eag{} and \tng{} simulations, with maximum quenching excesses factor (for the fiducial matching scheme) of $11$ and $2$, respectively. 
    In contrast, \ill{} demonstrates a quenching deficit, such that more control galaxies quench compared with the post-merger population (see Figure~\ref{fig:excess}).
    
    \medskip
    
    \item We proved the lack of a rapid merging-driven quenching scenario.
    However, we cannot statistically rule out a possible long-term quenching (over a few Gyr after coalescence) due, for instance, to a prolonged cooling time of the circumgalactic medium.
    For such long timescales we find that in all three simulations quenching in the post-merger population is statistically indistinguishable from the control sample of non-interacting galaxies, thus preventing any attempt to causally connect long quenching to merger itself (see figures~\ref{fig:fractions} and \ref{fig:excess}, and comments in Section~\ref{sec:rapid_vs_long}).
    
    \medskip
    
    \item We compare our results with the observational results of \citet{Ellison2022}, looking at the quenched fraction and quenching excess in post-starburst galaxies. We find the quenched fractions of \eag{} and \ill{} to be almost comparable to the lower limit estimated from E+A post-starburst selection, but incompatible with the lower limit estimated from PCA post-starburst selection. We further find that the simulations generally fail to reproduce the quenching excess of \citet{Ellison2022}, with a note that \eag{} is able to produce an excess in agreement with \citet{Ellison2022} but with poor statistical significance (see Section~\ref{sec:observations}, and Figure~\ref{fig:observation}).
    
    \medskip
    
    \item We investigate the rarity of quenching in the post-merger population and find that the majority of post-mergers retain a significant amount of gas in the 500 Myrs following coalescence, providing sufficient fuel for sustaining star formation (see Section~\ref{sec:discussion}, and Figures~\ref{fig:eag_analysis}, \ref{fig:Ill_analysis}, and \ref{fig:tng_analysis}). 
    
    \medskip
    
    \item We further demonstrate that there are two mechanisms at play to quench post-mergers: (1) a predisposition for post-mergers to quench if they exhibit a low gas fractions at the time of coalescence, and (2) a large amount of gas ejected promptly within 500 Myrs of coalescence (see Section~\ref{sec:gas_analysis}, and figures~\ref{fig:gas_loss} and~\ref{fig:gas_loss_hist}).

\end{itemize}

\section*{Acknowledgements}
%We acknowledge and respect the lək̓ʷəŋən peoples on whose
%traditional territory the university stands and the Songhees, Esquimalt and W̱SÁNEĆ peoples whose historical relationships with the land continue to this day.
We thank the referee, Neven Caplar, for his useful comments.
We would like to express our gratitude to Jonathan J. Davies for his constructive comments and suggestions that greatly contributed to this work and Stuart McAlpine for providing the private-available \eag{} snipshots galaxy catalogue and insightful discussions. We are also grateful to Andrew Pontzen for his useful comments and suggestions.  
DRP and SLE gratefully acknowledge NSERC for
Discovery Grants which helped to fund this research.  
This research was enabled, in part, by the
computing resources provided by WestGrid and Compute Canada.

%%%%%%%%%%%%%%%%%%%%%%%%%%%%%%%%%%%%%%%%%%%%%%%%%%
\section*{Data Availability}

The data used in this work have been previously published. 
\tng{} and \ill{} data are publicly available at \url{https://www.tng-project.org}, and \url{www.illustris-project.org/}, respectively.
\eag{} data, instead, can be accessed via \url{https://virgodb.dur.ac.uk}, however the snipshots used in this paper are for internal use only.



\bibliographystyle{mnras}
\bibliography{Bibliography_v000} % if your bibtex file is called example.bib

%\begin{thebibliography}{99}
%\end{thebibliography}

%%%%%%%%%%%%%%%%%%%%%%%%%%%%%%%%%%%%%%%%%%%%%%%%%%

%%%%%%%%%%%%%%%%% APPENDICES %%%%%%%%%%%%%%%%%%%%%

\appendix

\bsp	% typesetting comment
\label{lastpage}
\end{document}